\documentclass{aa}
\authorrunning{M. Volonteri}
\usepackage{psfig}
\usepackage{amssymb}
\begin{document}

   \thesaurus{06     % A&A Section 6: Form. struct. and evolut. of stars
              (03.11.1;  % Cosmogony,
               16.06.1;  % Planets and satellites: general,
               19.06.1;  % Solar system: general,
               19.37.1;  % Stars: formation of,
               19.53.1;  % Stars: oscillations of,
               19.63.1)} % Stars: structure of.
   \title{Interpreting the optical data of the Hubble Deep Field South: colors,
 morphological number counts and photometric redshifts}

   \author{ M. Volonteri$^{2}$, P. Saracco$^1$, G. Chincarini$^{1,2}$ and M. Bolzonella$^{2,3}$}

   \offprints{Marta Volonteri e-mail marta@merate.mi.astro.it}

    \institute{$^1$ Osservatorio Astronomico di Brera-Merate, Italy, $^2$ Dipartimento di Fisica, Univ. di Milano, Italy, $^3$ Istituto di Fisica Cosmica ``G.Occhialini'', Milano, Italy } 
   \date{Received ......, 2000 ; accepted }
\titlerunning{Interpreting the optical data of the Hubble Deep Field South}
     \maketitle

\begin{abstract}
We present an analysis of the optical data of the Hubble Deep Field
South. We derive F300W$_{AB}$, F450W$_{AB}$, F606W$_{AB}$ and
F814W$_{AB}$ number ~counts ~for ~galaxies ~in ~our ~catalogue
(Volonteri et al. 2000): the number-counts relation has an increasing
slope up to the limits of the survey in all four bands.  The slope is
steeper at shortest wavelengths: we estimated
$\gamma_{F300W_{AB}}\sim0.47\pm0.05$,
$\gamma_{F450W_{AB}}\sim0.35\pm0.02$,
$\gamma_{F606W_{AB}}\sim0.28\pm0.01$ and
$\gamma_{F814W_{AB}}\sim0.28\pm0.01$.  The color-magnitude relations
of galaxies shows an initial blueing trend, which gets flat in the
faintest magnitude bin and the sample contains a high fraction of
galaxies bluer than local sources, with about $50\%$ of sources with
F814W$_{AB}>27$ having (F450W-F606W)$_{AB}$ bluer than a typical local
irregular galaxy.  Morphological number counts are actually dominated
by late type galaxies, while early type galaxies show a decreasing
slope at faint magnitudes.  Combining this information with
photometric redshifts, we notice that galaxies contributing with a
steep slope to the number counts have $z\gtrsim1$, suggesting a
moderate merging.  However we emphasize that any cut in apparent
magnitude at optical wavelengths results in samples biased against
elliptical galaxies, affecting as a consequence the redshift
distributions and the implications on the evolution of galaxies along
the Hubble sequence.
                    
\end{abstract}

%
%________________________________________________________________

\section{Introduction}

The Hubble Deep Field South (HDF-S) was observed in October 1998 by
the Hubble Space Telescope (HST).  It is the southern counterpart of
the Hubble Deep Field North (HDF-N) and shares its characteristics of
depth and spatial resolution.  The HDF-S is a four arcmin$^2$ survey
located at RA 22$h$ 32$m$ 56$s$, DEC -60$^\circ$ 33' 02'', observed
during 150 orbits with the Wide Field Planetary Camera 2
(WFPC2). HDF-S images cover a wavelength range from the ultraviolet to
near-infrared with 4 broad-band filters: F300W, F450W, F606W,
F814W.  Details about observations and data reduction may be found in
the~ Hubble~ Deep~ Field~ South~ web page (http://www.stsci.edu/ ftp/science/hdfsouth/hdfs.html).

The main goal of these surveys, HDF-N and HDF-S, is to study faint
galaxies, in particular to investigate the nature of the faint blue
galaxies excess and the evolution of galaxies (e.g. Ellis 1997).

Faint galaxy counts have been of paramount importance to show
that galaxies do evolve with redshift, even if the overall scenario
and the physical processes that led to evolution are still debated. An
excess of observed galaxies with respect to a no-evolution model
begins to appear at B$\sim21-22$ (e.g. Maddox et al. 1990; Jones et
al. 1991; Metcalfe et al. 1991) and continues to rise up to B$>28$
(Metcalfe et al. 1995) exceeding by a factor 4 to 10, depending on the
choice of $q_0$ (0-0.5) and on the LF (see e.g. Koo \& Kron 1992;
Ellis 1997). This excess has actually undoubtly proved that galaxies
evolve. However, the analysis of the optical data have not offered a
unequivocal interpretation: the steepening of the faint end luminosity
function (LF) of galaxies at high redshifts (Lilly et al. 1995; Ellis
et al. 1996) suggests that either the typical luminosity of galaxies
was brighter by $\sim1$ magnitude at $z\sim0.7$ or that there were 2-3
times more galaxies at this epoch. Most likely both luminosity and
number density evolution are acting: luminosity evolution seems to
play an important role at least on disk galaxies as shown by the
results obtained by Schade et al. (1995) and Lilly et al. (1998) on
the CFRS-LDSS sample. On the other hand also mergers of galaxies seem
to significantly contribute to the evolution of both the luminosity
function and luminosity density of the Universe at least down to
$z\sim1$ (Le F\`evre et al. 2000).  Analyzing high resolution ground
based imaging of blue galaxies at redshift $z<0.7$, Colless et
al. (1994) actually noticed that most of them have a close companion
at a projected distance of about 10 kpc. This result suggests the
presence of merging at $z<0.5$, which could increase the star
formation, explaining the blue colors of the sources.

Disentangling the effects of luminosity and number density
evolution is fundamental in understanding the nature of field galaxy
formation and evolution, allowing a direct comparison with models. For
instance, semi-analytical models of galaxy formation in hierarchical
clustering scenarios, assuming that galaxies assemble through
successive merger events from smaller sub-units, make concrete
predictions about the evolution of merger rate, morphological mix and
relative redshift distribution (e.g. Baugh et al. 1996; Kauffmann
1996; Kauffmann \& Charlot 1998). In this scenario the number density
of galaxies should increase with redshift, their sizes should decrease
and the fraction of high redshift galaxies ($z>2$) in an IR-selected
sample should be very low. Moreover the morphological mix should
change in favour of late type and irregular galaxies which thus would
dominate both galaxy counts at faint magnitudes and the high redshift
tail of the redshift distribution, where a lack of elliptical galaxies
should be observed. Most of these predictions have been the object of
detailed analysis of data resulting in conflicting results. For
instance, some works claim a deficit of ellipticals at $z>1$ (Zepf
1997; Franceschini et al. 1998; Barger et al. 1999) and a very low
fraction ($\sim5\%$) of $z>2$ galaxies in a K$<21$ selected sample
(Fontana et al. 1999). On the contrary other authors find a constant
comoving density of ellipticals down to $z\sim2$  (e.g. Ben\'{\i}tez
et al. 1999; Broadhurst \& Bowens 2000) and a strong
clustering signal of EROs in K-selected samples (Daddi et
al. 2000). HST Medium Deep Survey (MDS), on the other hand, indicates
that a large fraction of galaxies at moderate redshift ($z\sim0.5$)
are actually intrinsically faint and exhibit peculiar morphologies
suggestive of merging (Griffiths et al. 1994; Driver et al. 1995;
Glazebrook et al. 1995; Neuschaefer et al. 1997). The results derived
from the morphological analysis on the HDF-N data (van den Bergh et
al. 1996; Abraham et al. 1996; Driver et al. 1998) seem to confirm the
high fraction of apparent irregular/peculiar galaxies with respect to
the local universe which may be the cause of the faint blue galaxies
excess (Ellis 1997).

Williams et al. (1996) stress also that HDF-N galaxies show a strong
bluing color trend: while red galaxies are compatible with an
elliptical galaxy colors, most of blue sources are bluer than an
irregular galaxy, and these colors might be accounted for only
considering luminosity evolution.

Thus it seems that the overall evolutionary scenario is still missed
and the understanding of the contribution of the different populations
of galaxies to this scenario rather confused.

In this work we present an analysis of number counts, colors and
morphological distributions of HDF-S galaxies in an attempt to reveal
and better constraint the evolving populations of galaxies.  The plan
of the paper is as follows: in Section 2 we summarize the steps
leading to the creation of the catalogue, the computation of optical
galaxy counts is explained in Section 3, in Section 4 the technique
used to determine galaxy colors. In Section 5 we discuss our results,
both counts and colors, while in Section 6 we present and discuss
morphological number counts and in Section 7, by adding the
information given by photometric redshifts and number counts models,
we disentangle the contribution of different galaxy populations and
discuss the interpretation of redshift distributions.  Finally, in
Section 8 we summarize our results and express our conclusions.

\section{Image Analysis}

The samples here used are not extracted from the public HDF-S catalogs
available on the Web.  On the contrary they have been independently
created beginning from the optical HST images in order to have a
control of the detection procedure and of magnitude estimate.  A
detailed description of the procedure adopted and the selection
criteria used to obtain the catalogs in the different bands, as well
as magnitude estimates, can be found in Volonteri et al. (2000). Here
we briefly describe the steps related to the derivation of number
counts.

The object detection was performed across the area observed in optical
bands (WFPC2 field) using SExtractor (Bertin and Arnouts 1996).
Images were first smoothed with a Gaussian function having FWHM equal
to the one measured on the images ($\approx$ 0.2 arcsec).  A
detection threshold of 1 sigma per pixel and a minimum detection area
equal to the seeing disk ($\approx 0.02$ arcsec$^2 = 13$
pixel) were adopted to peak up objects.  This threshold corresponds
to a minimum signal-to-noise ratio of $S/N_{WF}=1.34$ and
$S/N_{PC}=0.67$ for the faintest sources detectable on the WF area and
on the PC area respectively.

In Table 1 we report for each filter the zero-point (AB magnitude,
Oke 1974), the sky RMS estimated by SExtractor and the corresponding
5$\sigma$ limiting magnitude for a point source.
\begin{table}
\caption{RMS sky values and 5$\sigma$ magnitude limit for
the different bands.}
\begin{center}
\begin{tabular}{llll} 
\hline
Filter& zero-point& RMS&m$_{lim}$  \\
 &   &  (ADU/pix)$\times10^{-5}$& \\
\hline
F300W & 20.77&	1.674& 28.87\\
F450W& 21.94&	2.284&	29.71\\
F606W&	23.04&	4.126&	30.16\\
F814W&	22.09&	2.960&	29.58\\
\hline
\hline
\end{tabular} 
\end{center}
\end{table}

``Pseudo-total'' magnitudes were estimated using the method of
Djorgovski et al. (1995) and Smail et al. (1995) on the basis of the
following steps:
\begin{itemize}
\item the SExtractor isophotal corrected magnitude has been assigned
to \emph{large} sources, i.e. those sources having an isophotal
diameter $D_{iso}>\theta_1$ and to those sources flagged by SExtractor
as ``blended'';
\item the aperture corrected magnitude (estimated within $\theta_1$
and then corrected to $\theta_2$, being $\theta_1<\theta_2$) has been
assigned to \emph{small} sources with $D_{iso}<\theta_1$;
\end{itemize}
$\theta_1$ has been defined as the minimum apparent diameter of a
galaxy having an effective diameter $r_e=10$ kpc. Hereafter we
use a $\Lambda=0$ cosmology, with $q_0=0.5$ and $H_0=50$ km
s$^{-1}$Mpc$^{-1}$ unless differently specified. With this choice
$\theta_1=1.2$ arcsec.  $\theta_2$ is the diameter corresponding to
the area for which 90$\%$ of the bright sources have a smaller
isophotal area (see Volonteri et al. 2000 for a complete
description of procedure, tests and results).  
This procedure was applied independently for each band.

\begin{figure}
\centerline{\psfig{figure=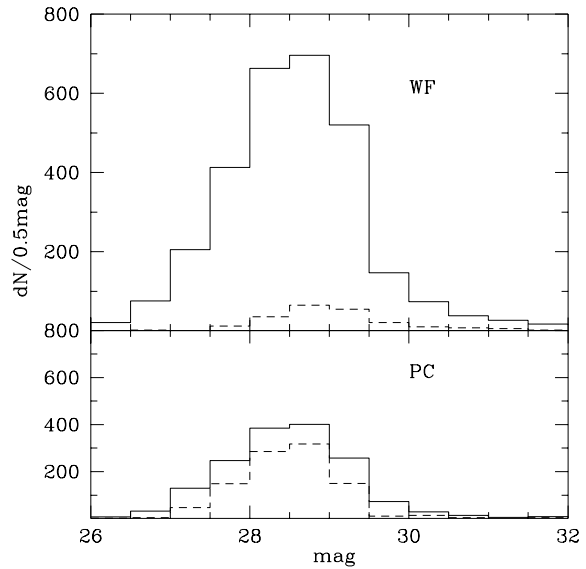,height=80mm}}
\caption{Spurious sources, defined as detections on the reversed (i.e.
multiplied by -1) F606W image.  The upper panel refers to the
WF field, the lower panel to the PC field. The solid line represents
the initial detections, the dashed line represents spurious sources
left after applying our selection criteria (i.e. a S/N=5 cut
off, after removing the edges of the images).}
\label{spurie}
\end{figure}

A measure of the detection reliability is necessary in order to evaluate
the number of spurious sources included in the sample.
%Since the completeness level is a function of object magnitude and size,
%our test will provide a guess for the optimum S/N ratio to extract our 
%fiducial samples.
We treated this problem statistically, in the hypothesis that
noise is symmetrical with respect to the mean sky value. Operationally
we first created for each filter a noise frame by reversing the
original images to reveal the negative fluctuations and to make
negative (i.e. undetectable) real sources (Saracco et al. 1999). 
Then we run SExtractor with the same detection parameter set used to
search for sources in the original images detecting, by definition,
only spurious sources.  In Figure~\ref{spurie} the magnitude
distribution of spurious sources obtained on the WF area and the PC
area in the F606W band are shown respectively.  Applying a S/N=5 cut
off is sufficient to reduce the spurious contamination to a negligible
fraction (4$\%$) on the WF area, after removing the edges of the
images with lower sensitivity.  On the contrary such a cut off is not
able to reduce spurious detections to a reasonable level on the PC
area, contamination being more than 35$\%$.  Thus due to such a large
number of spurious sources which would be introduced by the PC data,
we restricted the selection of sources to the central WF area only
corresponding to 4.38 arcmin$^2$.  On this area 450, 1153, 1694 and
1416 sources were selected accordingly to the above criteria in the
F300W, F450W, F606W and F814W band
respectively.

\section{Optical Galaxy Counts}

To derive galaxy counts, we first removed stars from the sample by
using the SExtractor star$/$galaxy classifier.  We defined as stars
those sources brighter than F814W$_{AB}=22$ and having a value of the
SExtractor ``stellarity'' index larger than 0.9.  This choice will
tend to underestimate stars both at faint magnitudes where no
classification is considered, and at bright magnitudes where some
fuzzy stars could be misclassified as galaxies.  On the other hand
this will ensure that our galaxy sample is not biased against compact
galaxies.  The star ``cleaning'' procedure has classified and removed
14 stars with F814W$_{AB}<22$ in agreement with the number of stars
found in the HDF-N by Mendez et al. (1998) to this depth and in excess
by a factor of two with respect to the prediction of the galaxy model
of Bahcall \& Soneira (1981).

\begin{table*}
\caption{Differential number counts in F300W$_{AB}$, F450W$_{AB}$,
F606W$_{AB}$ and F814W$_{AB}$ bands over the inner 4.38 arcmin$^2$ of
the WF area.}
\begin{center}
\begin{tabular}{rrlrrrrlrr} 
\hline
 F300W$_{AB}$ & $n_r$  & $\bar c$& $N/$0.5 mag/deg$^2$ & $\sigma_N$& F450W$_{AB}$ & $n_r$  & $\bar c$& $N$/0.5 mag/deg$^2$ & $\sigma_N$ \\
\hline
22.25 &--- &--- &--- &--- & 22.25 & 3 & 1& 2470 & 1625\\	
22.75 & 1 &1 & 820 & 850 &22.75 & 6 &1 & 4900 & 2405\\
23.25 & 1 &1 & 820 & 860 &23.25 & 8 &1 & 6575 & 2754 \\
23.75 & 7 &1 & 5750 & 2510&23.75 & 22 &1 & 18080 & 5152  \\
24.25 & 14 &1 & 11510 & 3730&24.25 & 27 &1 & 22190 & 5541 \\
24.75 & 20 &1 & 16440 & 4466&24.75 & 26 &1 & 21370 & 5074 \\
25.25 & 24 &1 & 19730 & 4772&25.25 & 80 &1 & 65750 & 10434 \\
25.75 & 50 &1 & 41100 & 7345&25.75 & 83 &1 & 68220 & 9892  \\
26.25 & 93 &1 & 76440 & 10594& 26.25 & 121 &1 & 99450 & 12025 \\
26.75 & 103 & 1.73  &179470 & 21455&26.75 & 153 &1 & 125750 & 13211 \\
27.25 & 286 & 4.42 & 767930 & 93729&27.25 & 220 &1 & 200710 & 22752  \\
27.75 &--- &--- &--- &--- & 27.75 & 208 &1.56  & 266700 & 24395 \\
28.25 &--- &--- &--- &--- &	28.25 & 137 &3.42 & 365960 & 35267\\ 		 	
\hline
\hline

F606W$_{AB}$ & $n_r$  & $\bar c$ & $N$/0.5 mag/deg$^2$  & $\sigma_N$ & F814W$_{AB}$ & $n_r$  & $\bar c$ & $N$/0.5 mag/deg$^2$  & $\sigma_N$ \\  
\hline 
22.25 & 7 & 1& 5753 & 2764&22.25 & 11 & 1& 8630 & 3397\\
22.75 & 13 &1 & 10680 & 3986&22.75 & 28 &1 & 21960 & 6169\\
23.25 & 15 &1 & 12330 & 4102&23.25 & 26 &1 & 20390 & 5646 \\
23.75 & 27 &1 & 22190 & 5797& 23.75 & 35 &1 & 27450 & 5533\\
24.25 & 28 &1 & 23010 & 5533& 24.25 & 37 &1 & 29020 & 6071\\
24.75 & 38 &1 & 31230 & 6397&  24.75 & 65 &1 & 50980 & 8032\\
25.25 & 72 &1 & 59180 & 9346& 25.25 & 94 &1 & 73720 & 9878\\
25.75 & 90 &1 & 73970 & 10183& 25.75 & 139 &1 & 109020 & 11794\\
26.25 & 129 &1 & 106030 & 12228&26.25 & 173 &1 & 135690 & 12949 \\
26.75 & 163 &1 & 133970 & 13440&26.75 & 235 &1 & 184310 & 14801 \\
27.25 & 202 &1 & 166027 & 14600& 27.25 & 342 &1.08 & 289700 & 23298\\
27.75 & 283 &1  & 232602 & 17254&27.75 & 332 &1.86  & 484330 & 32485 \\
28.25 & 313 &1.31 & 344730 & 26643&--- &--- &--- &--- &--- \\
28.75 & 190 &2.84 & 416960 & 36212&--- &--- &--- &--- &--- \\
\hline
\hline

\end{tabular} 
\end{center}
\end{table*}

Completeness correction for faint undetected sources strongly depends
on the source apparent spatial structure and on their magnitude.  The
high resolution of HDF-S images allows the detection of sub-galactic
structures, such as HII regions.  Moreover the light observed from a
high fraction of the galaxies in the HDF-S is emitted in the UV and
F450W pass-bands, so that the observed morphology is strongly
affected by star formation episodes.  These features imply that
``typical'' profiles of galaxies are not able to well describe the
shapes of a lot of galaxies in the HDF-S.  Thus, in order to reproduce
the manifold of shapes which characterizes sources in the HDF-S,
following Saracco et al. (1999), we generated a set of simulated
frames by directly dimming the original frames by various factors
while keeping constant the RMS.  This procedure has allowed us to
avoid any assumption on the source profile providing an artificial
fair dimmed sample in a real background noise even if it cannot take
into account differences due to evolution.  We thus define the
correction factor $\bar c$ as the mean number of dimmed galaxies which
should enter the fainter magnitude bin over the mean number of
detected ones.  In Table. 2 we report the raw counts $n_r$, the
completeness correction factor $\bar c$, the counts per square degree
corrected for incompleteness $N$ and their errors $\sigma_N$.

Colley et al. (1996) suggested that galaxies in the HDF-N may
suffer from a wrong selection. High redshift galaxies on optical
images have a lumpy appearance: first the redshift moves the
ultraviolet rest-frame light into the optical, so~ galaxies~ are~
observed~ in UV rest-frame, where star-forming regions are more
prominent; second the fraction of irregular galaxies is higher than
locally (van den Bergh et al. 1996, Abraham et al. 1996) and a large
number of galaxies may display asymmetry and multiple structure. We
treated this feature analyzing our sample in F814W and F450W. The
F814W-band catalogue should suffer less from the effects described
above, being selected in the reddest filter, the vice versa is true
for the F450W-band catalogue (but see Volonteri et al. 2000 for a more
detailed analysis). About 20-30$\%$ of sources in F450W-band catalogue
have separation $<1$ arcsec. We therefore analyzed these sources, by
cross-correlating F814W and F450W catalogues. In the F450W-band
catalogue we selected pairs with a separation $<1$ arcsec which were
not included in the F814W-band catalogue. These objects were single
sources split in the F450W-band (with F450W$_{AB}\approx $ 27-29),
corresponding to a single detection in the F814W-band. We then
used SExtractor on the F450W frame, after choosing a higher {\sc
deblend\_mincont}$=0.1$.

\begin{figure}
\centerline{\psfig{figure=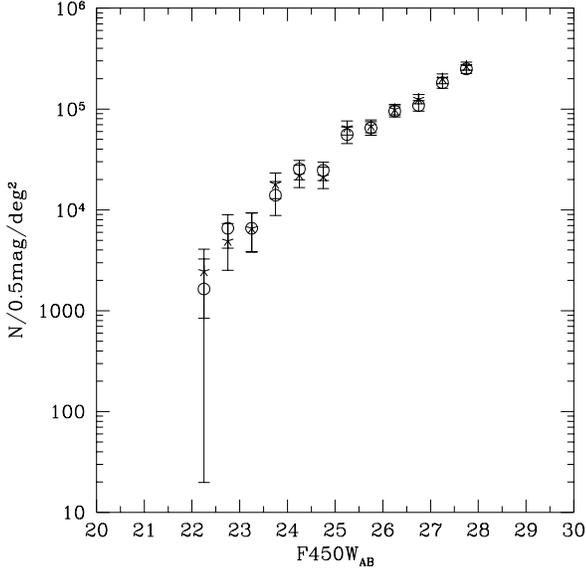,height=80mm}}
\caption{Comparison of F450W-band number counts uncorrected (stars) or
corrected (empty circles) for ``oversplitting'' of sources. The
faintest bin are unaffected by the correction, since it is within the
error. }
\label{BB}
\end{figure}

87 sources, with $21<$F450W$_{AB}<26$, corresponding
to about 7$\%$ of the whole sample, were then considered as single
galaxies. In Figure \ref{BB} counts obtained with the
uncorrected catalogue are shown as stars, while corrected counts are
shown as empty circles. In the faintest bins the correction is within
the error, accounting for less than 10$\%$ of the counts.

Errors were obtained by quadratically summing the Poissonian
contribution $\sigma_{n_r}=\sqrt{n_r}$ of raw counts, the contribution
due to clustering fluctuation
\begin{equation}
\sigma_\omega\sim\omega(\theta)^{1/2} N
\end{equation} 
being $\omega(\theta)=A_\omega\theta^{-(\gamma-1)}$ the angular correlation 
function, and the uncertainty on the correction factor 
\begin{equation}
\sigma_{\bar c}=\sqrt{1/k\sum(c_i-\bar c)^2}
\end{equation}
where $k$ is the number of frames dimmed for each dimming factor.
Assuming that the amplitude $A_\omega$ evolves with magnitude
on the basis of the relation log$A_\omega=-0.3\times mag+const$ 
(Brainerd et al. 1994; Roche et al. 1996) we derived
\[\log A_\omega(F300W_{AB})=4.438-0.3 F300W_{AB}\] 
\[\log A_\omega(F450W_{AB})=4.314-0.3F450W_{AB}\] 
\[\log A_\omega(F606W_{AB})=4.438-0.3F606W_{AB}\] 
\[\log A_\omega(F814W_{AB})=4.161-0.3 F814W_{AB}.\]
 
\begin{figure}
\psfig{figure=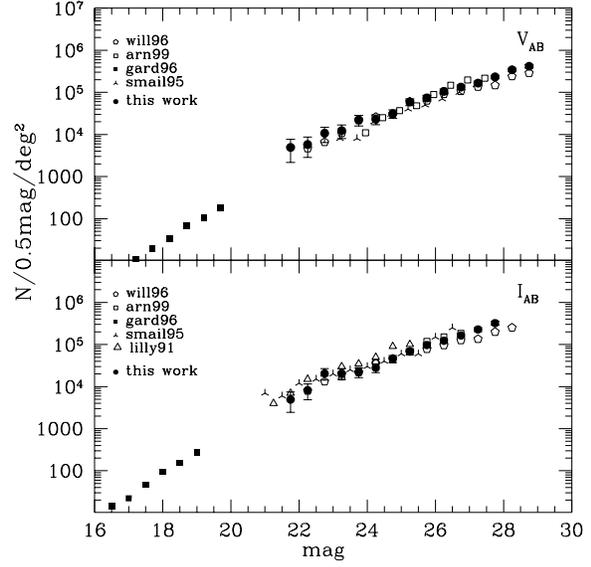,angle=0,width=8cm}
\caption{A comparison of number counts computed in this work with
counts obtained in literature in V$_{AB}$ and I$_{AB}$.
In particular Williams et al. (1996, will96) refer to the HDF-N, not
corrected for incompleteness. Other symbols correspond to Arnouts et
al. (1999, arn99), Gardner et al. (1996, gard96), Smail et al. (1995,
smail95), Lilly et al.  (1991, lilly91). Our best fit slopes are
$\gamma_{F606W_{AB}}=0.28$ and $\gamma_{F814W_{AB}}=0.28$, in
agreement with previous works, $d \log N/dm \approx 0.27-0.32$
(e.g. Smail et al.(1995), Arnouts et al. (1999)), while Williams et
al. (1996) find $d \log N/dm=0.17$ for F606W$_{AB}$=26-29, $d \log
N/dm=0.35$ for F606W$_{AB}$=23-26 and $d \log N/dm=0.18$ for
F814W$_{AB}$=26-29, $d \log N/dm=0.31$ for
F814W$_{AB}$=23-26.}
\label{countsVI}
\end{figure}

\begin{figure}
\centerline{\psfig{figure=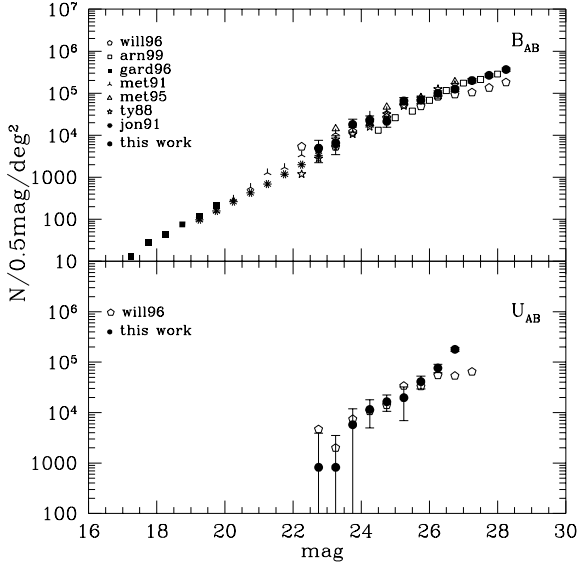,height=80mm}}
\caption{As in Figure \ref{countsVI}, in B$_{AB}$ and U$_{AB}$
bands.  The comparison with previous works includes Williams et al. on
the HDF-N (1996, will96), Arnouts et al. (1999, arn99), Gardner et
al. (1996, gard96), Metcalfe et al. (1991, met91), Tyson et al. (1988,
ty88), Jones et al.  (1991, jon91). Slopes are
$\gamma_{F450W_{AB}}$=0.35 e $\gamma_U=0.47$. Tyson et al.  (1988)
find $d \log N/dm=0.45$, Arnouts et al. (1999) $d \log N/dm=0.31$ in
the F450W$_{AB}$. In F300W$_{AB}$ our result is
consistent with $d \log N/dm \approx 0.4-0.6$ (Hogg et al. (1997), Koo
(1986)).  Williams et al. (1996), find $d \log N/dm=0.16$ for
F450W$_{AB}$=26-29, $d \log N/dm=0.39$ for F450W$_{AB}$=23-26 e $d
\log N/dm=0.05$ for F300W$_{AB}$=26-28, $d \log N/dm=0.40$ for
F300W$_{AB}$=23-26, Pozzetti et al.  (1998) find $d \log
N/dm=0.135$.}
\label{countsBU}
\end{figure}

The number counts here derived in the F300W and F450W
bands and in the F606W and F814W bands are shown in
Figures \ref{countsVI} and \ref{countsBU} together with those from the
literature. The relation between counts and magnitude may be written
as
\[\frac{d\log N}{dm}= \left\{\begin{array}{ll}
\gamma_1 m+c_1 & m \leq m'\\
\gamma_2 m+c_2 & m > m'\\
\end{array}
\right. \]
with $\gamma_1 > \gamma_2$. The knee $m'$ is at B$\approx$25 (Lilly et
al.  1991, Metcalfe et al. 1995), and the value of $\gamma_1$ varies
between 0.4 and 0.6 according to the band and analogously for
$\gamma_2$ it varies between 0.2 and 0.5. We estimated both $\gamma_1$
and $\gamma_2$ in F450W$_{AB}$, F606W$_{AB}$ and F814W$_{AB}$,
obtaining for
$\gamma_1$: $\gamma_{1,F450W_{AB}}\sim0.4\pm0.1$, 
$\gamma_{1,F606W_{AB}}\sim0.34\pm0.1$ and 
$\gamma_{1, F814W_{AB}}\sim0.62\pm0.1$, and for 
$\gamma_2$: $\gamma_{2,F450W_{AB}}\sim0.19\pm0.01$, 
$\gamma_{2,F606W_{AB}}\sim0.19\pm0.1$ and 
$\gamma_{2, F814W_{AB}}\sim0.19\pm0.1$. 

The most notable feature appears in the F300W-band counts: these are
described by a slope $\gamma_{F300W_{AB}}=0.47\pm0.05$ which is much
steeper than the value $\gamma_{F300W_{AB}}\sim0.15$ derived by
Williams et al. (1996) and Pozzetti et al. (1998) on the HDF-N data in
the same magnitude range.  Such a steep slope, which agrees with the
findings of Hogg et al. ($\gamma_U\sim0.5$; 1997) and Fontana et
al. ($\gamma_U\sim0.49$; 1999), does not depend on a possible
over-estimate of the incompleteness in the faintest magnitude bins:
the same slope is actually described by counts at magnitudes
U$_{300}<26$ where the sample is 100$\%$ complete.  Moreover the
F300W-band counts do not show evidence of any turnover or flattening
down to F300W$_{AB}$=27, contrary to what is claimed by Pozzetti et
al. (1998).
We will discuss this issue in Section 5.1.  The amplitude and slope of
the number counts in the whole range of magnitudes in the other
optical bands are in a good agreement with those previously derived by
other authors.  We estimated $\gamma_{F450W_{AB}}\sim0.35\pm0.02$,
$\gamma_{F606W_{AB}}\sim0.28\pm0.01$ and $\gamma_{
F814W_{AB}}\sim0.28\pm0.01$ for the F450W$_{AB}$, F606W$_{AB}$ and
F814W$_{AB}$ counts respectively.

\section{Optical Galaxy Colors}

In order to measure colors unbiased with respect to the selection
band, we created a \emph{metaimage} by summing all four frames, after
normalizing them to the same rms sky noise (see also Volonteri et
al. 2000). We run SExtractor in the so-called {\sc double image mode}:
detection and isophote boundaries were measured on the combined image,
while isophotal magnitudes were~ measured~ on~ F300W,
F450W, F606W, F814W images individually.  Using~
this~ procedure~ (Moustakas~ et~ al. 1997)~ both~ object~ detection~ and~
isophote determination are based on the summed image, and isophotes
are not biased towards any of the bands.

\begin{figure*}
{\centering\leavevmode
\psfig{figure=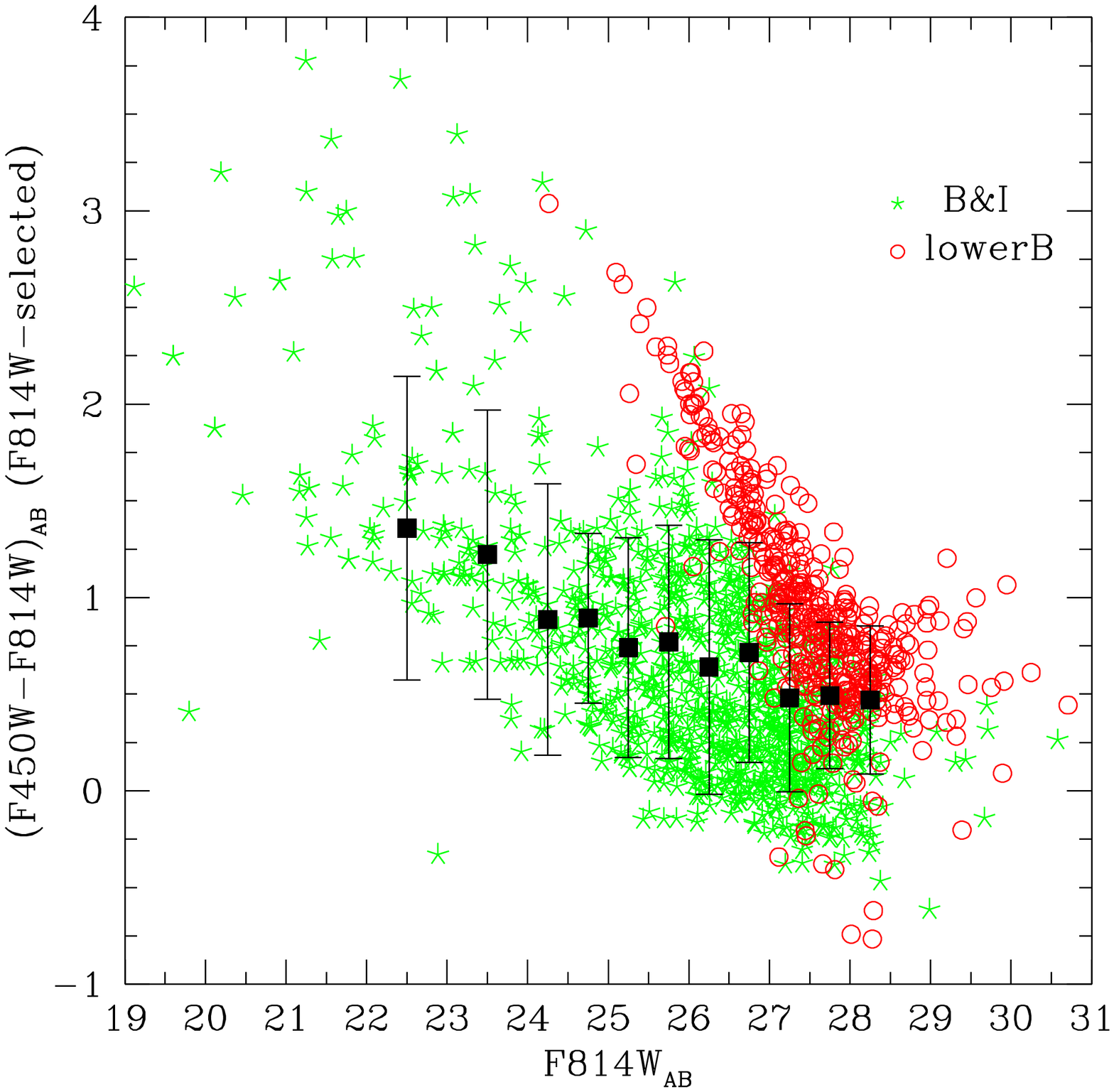,width=.49\textwidth}
\psfig{figure=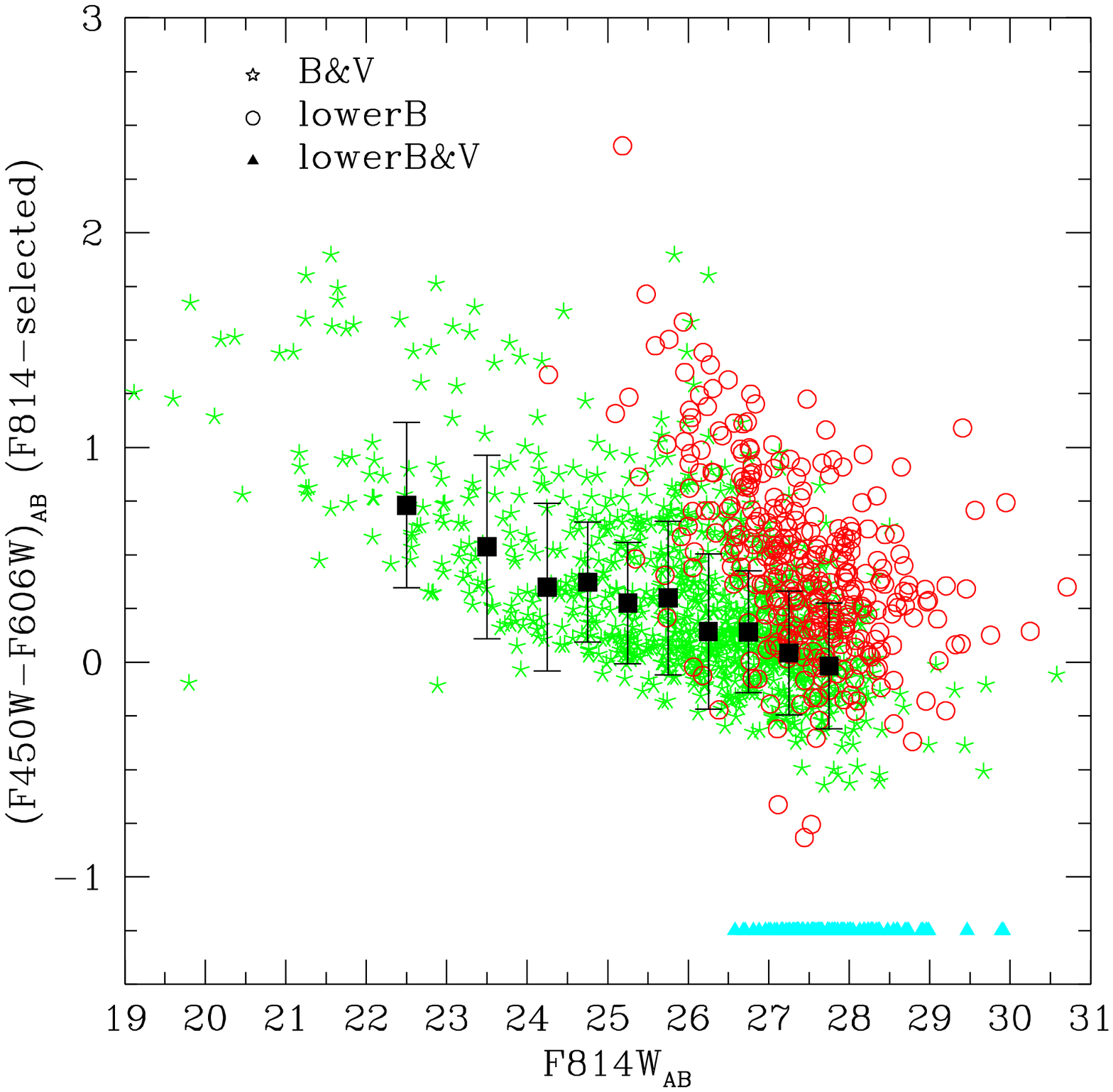,width=.49\textwidth}}
\caption{Left panel: (F450W$-$F814W)$_{AB}$ vs F814W$_{AB}$ for the F814W-selected
sample: in every magnitude bin the median colour is shown, error bars
are the standard deviation of the colour distribution. Asterisks
represent sources detected both in F450W and F814W
bands, while circles mark sources undetected in the F450W band:
their colour is a lower limit, defining the F450W$_{AB}$ magnitude as
the $5\sigma$ flux within the isophote on the combined UBVI
image. Right panel: (F450W$-$F606W)$_{AB}$   vs F814W$_{AB}$ for the F814W-selected sample:
asterisks represent sources detected in all F450W, F606W
and F814W bands, circles mark sources undetected in the
F450W band and triangles sources undetected both in
F450W and F606W bands (we assigned to those sources an
arbitrary value (F450W$-$F606W)$_{AB}=-1.25$). }
\label{colori_I1}
\end{figure*}

\begin{figure*}
{\centering\leavevmode
\psfig{figure=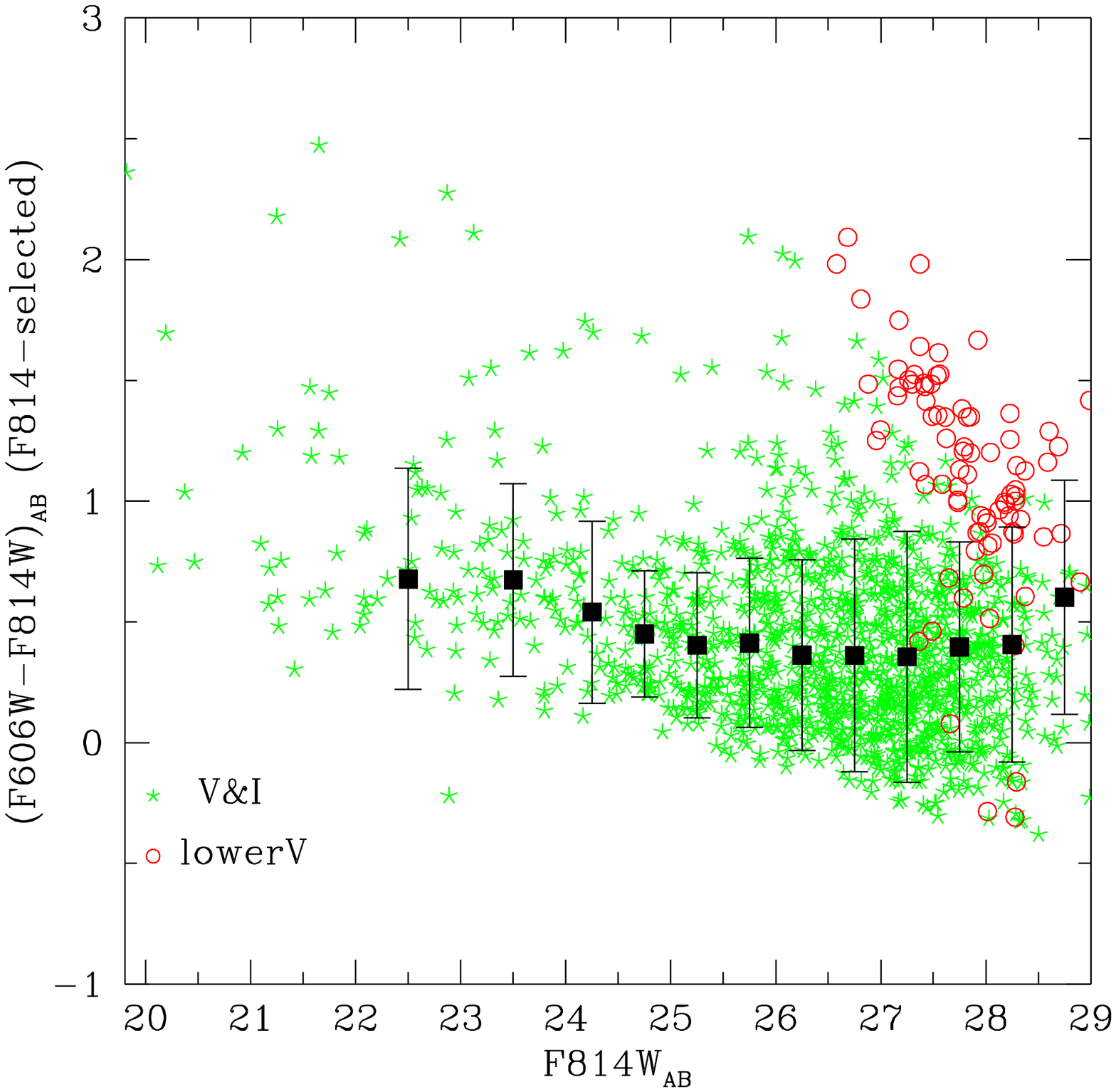,width=.49\textwidth}
\psfig{figure=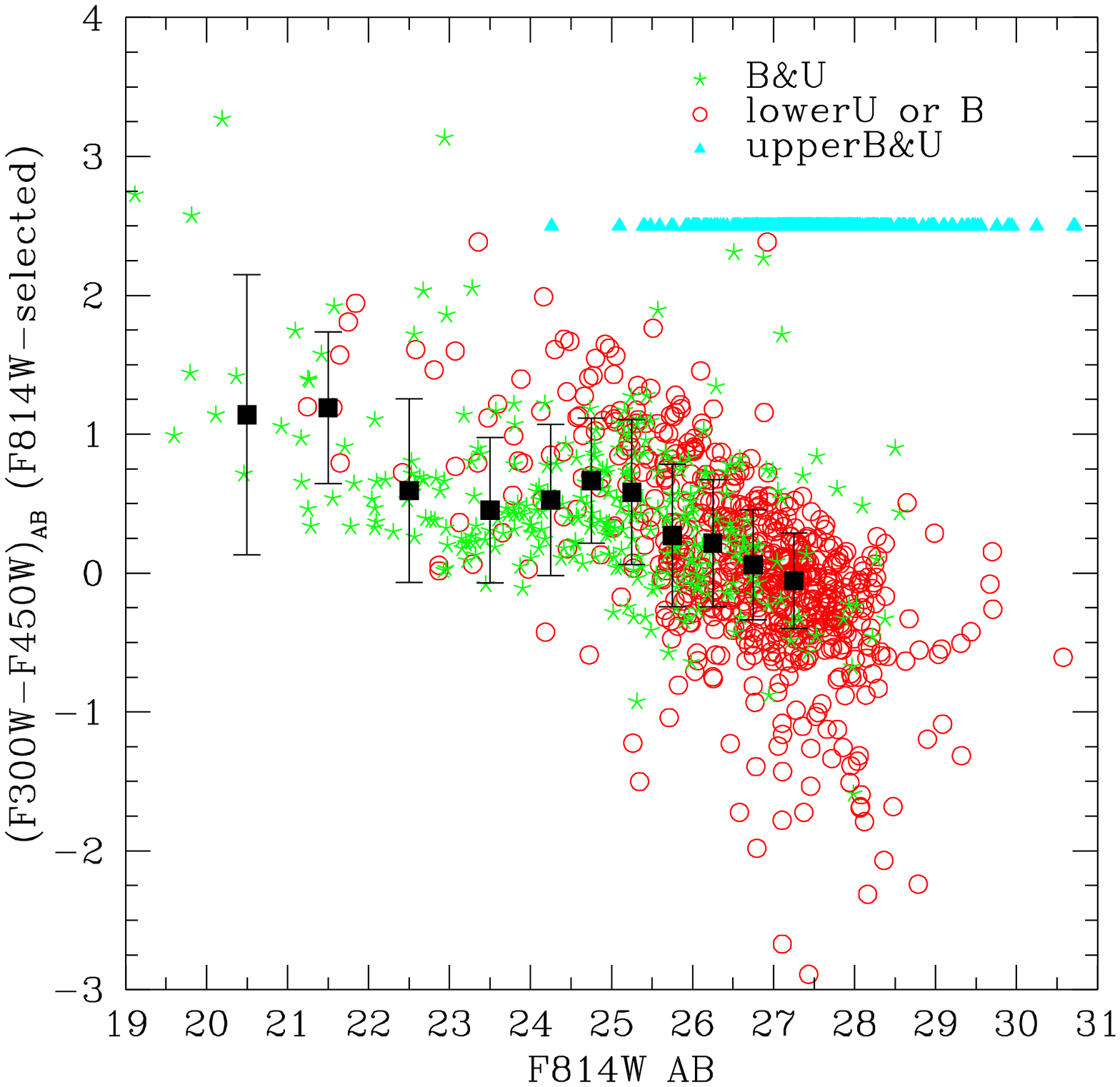,width=.49\textwidth}}
\caption{Left panel: (F606W$-$F814W)$_{AB}$ vs F814W$_{AB}$ for the F814W-selected
sample: asterisks represent sources detected in F606W and
F814W bands, circles mark sources undetected in the
F606W band. Right panel: (F300W$-$F450W)$_{AB}$ vs F814W$_{AB}$ for the
F814W-selected sample: asterisks represent sources detected in all
F300W, F450W and F814W bands, circles mark
sources undetected in the F300W band or F450W and
triangles sources undetected both in F450W  and F300W
bands (we assigned to those sources an arbitrary (F300W$-$F450W)$_{AB}=2.5$).}
\label{colori_I2}
\end{figure*}

\begin{figure*}
{\centering\leavevmode
\psfig{figure=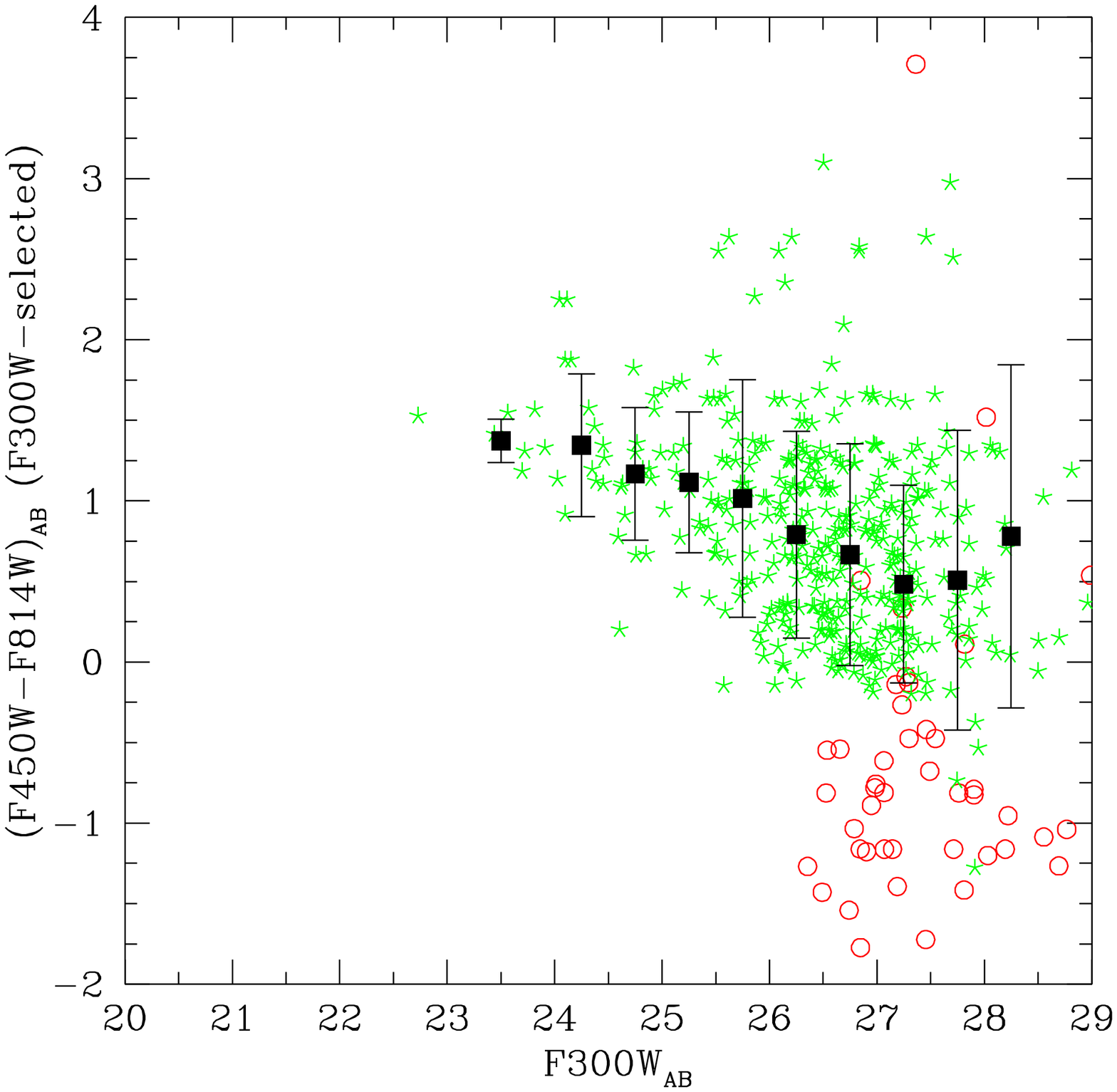,width=.49\textwidth}
\psfig{figure=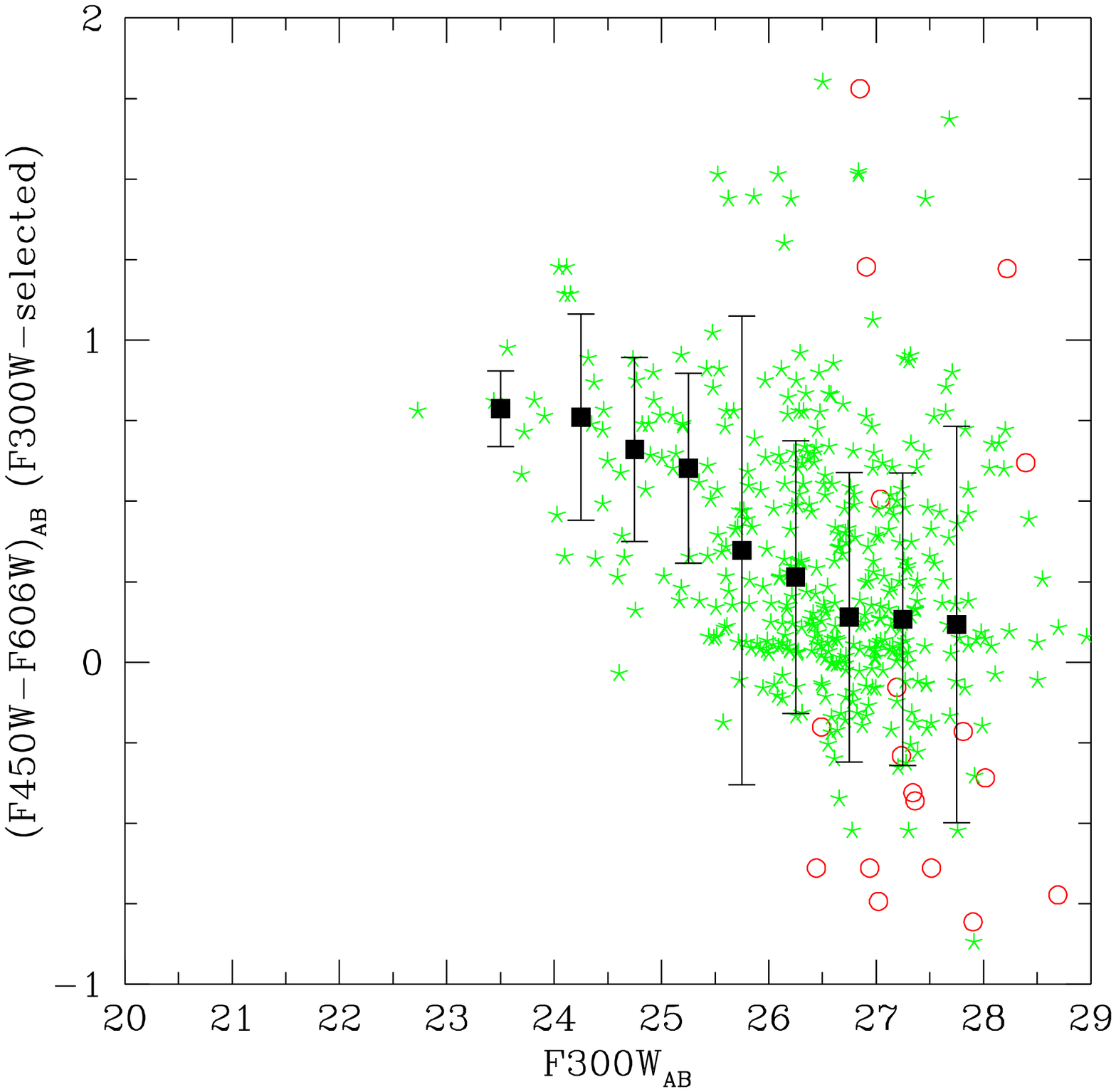,width=.49\textwidth}}
\caption{Left panel: (F450W$-$F814W)$_{AB}$  vs F300W$_{AB}$ for the
F300W-selected sample: in every magnitude bin the median colour
is shown, error bars are the standard deviation of the colour
distribution. Asterisks represent sources detected in F450W and
F814W and F300W bands, while circles mark sources
undetected in the F450W or F814W band. Right panel: (F450W$-$F606W)$_{AB}$
vs F300W$_{AB}$ for the F300W-selected sample. Asterisks
represent sources detected in F450W and F606W and
F300W bands, while circles mark sources undetected in the
F450W or F606W band.}
\label{colori_U1}
\end{figure*}

\begin{figure*}
{\centering\leavevmode
\psfig{figure=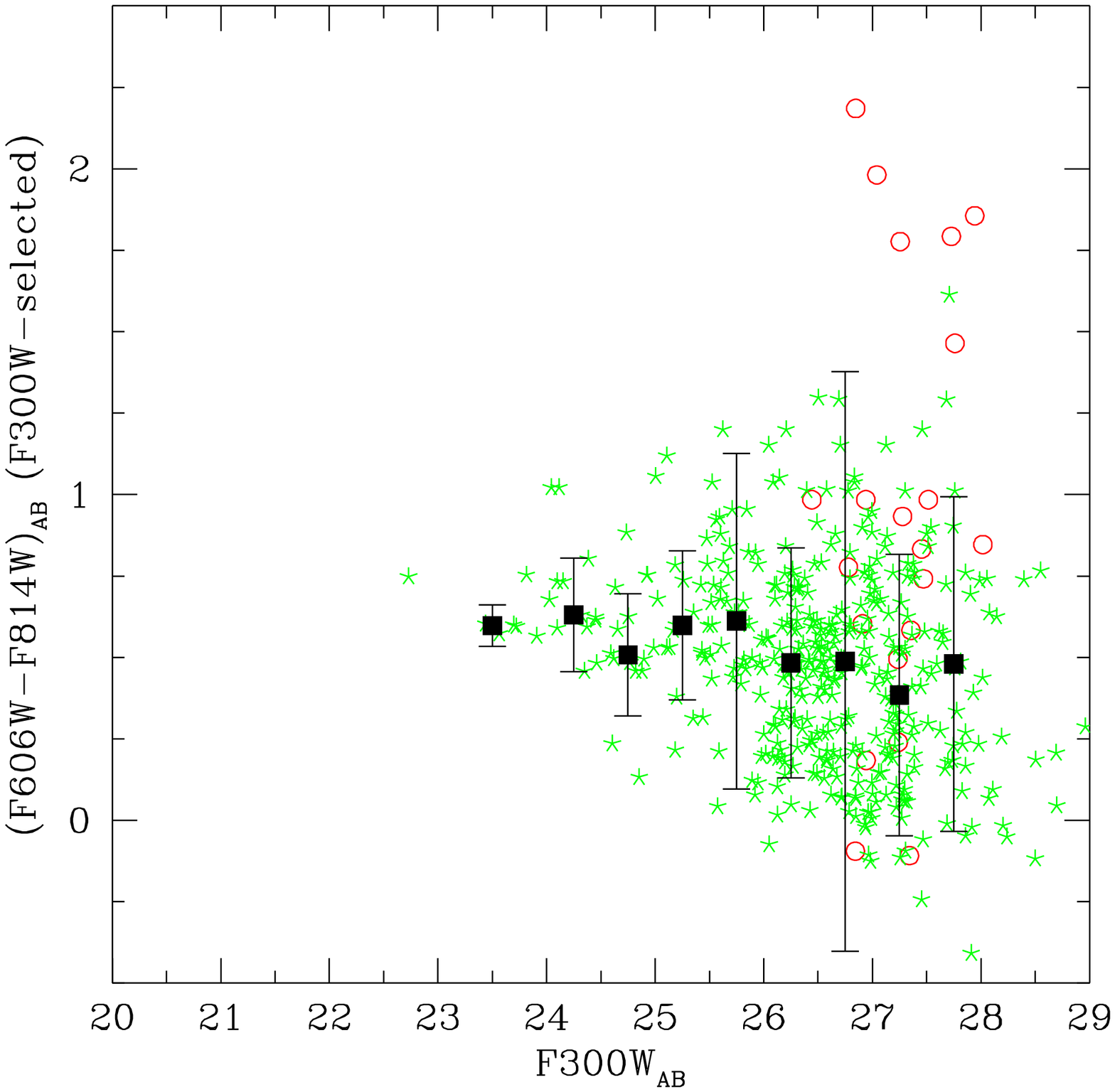,width=.49\textwidth}
\psfig{figure=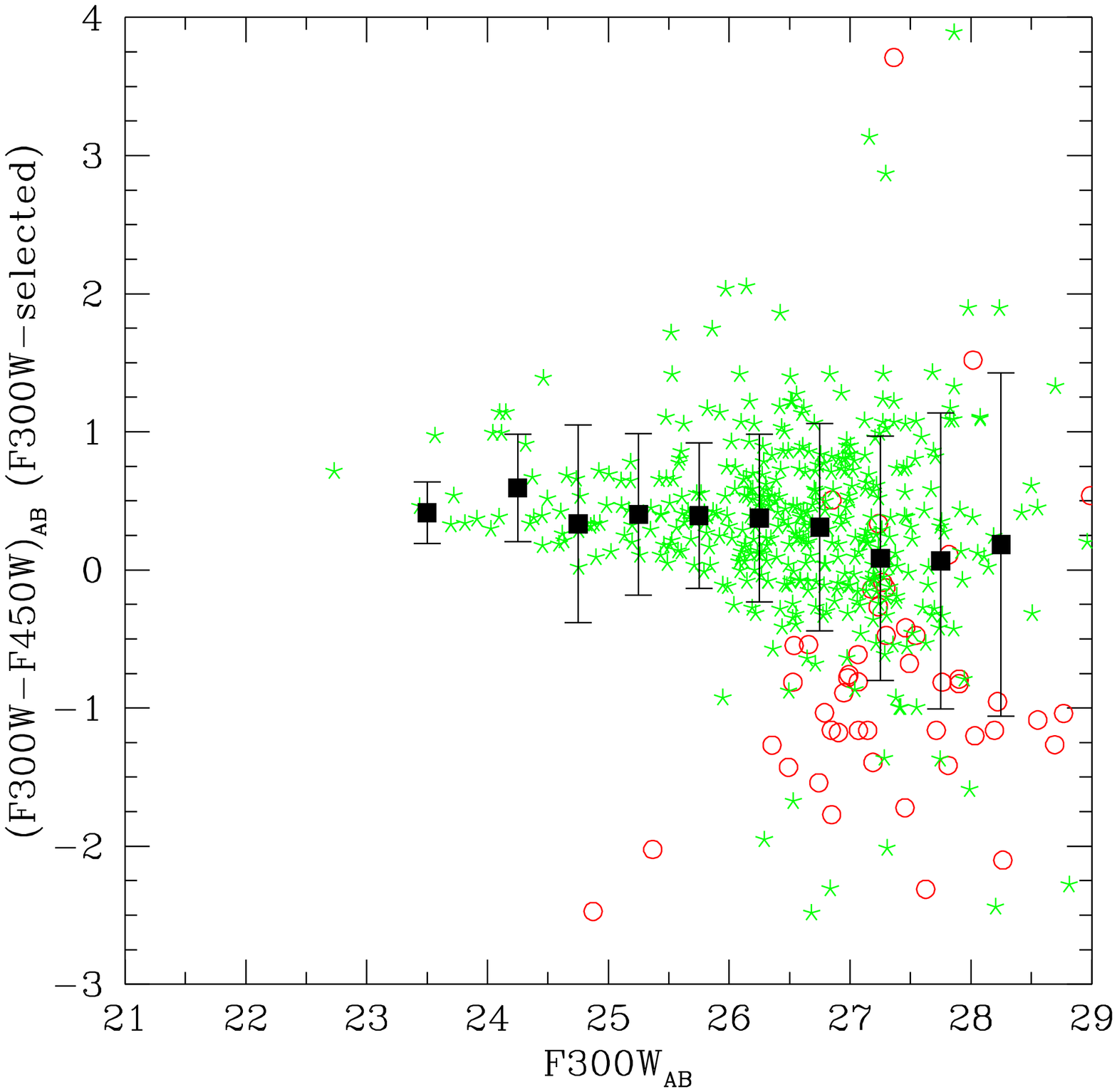,width=.49\textwidth}}
\caption{Left panel:  (F606W$-$F814W)$_{AB}$  vs F300W$_{AB}$ for the
F300W-selected sample.  Asterisks represent sources detected in
F606W and F814W and F300W bands, while circles mark
sources undetected in the F606W or F814W band. Right panel:
(F300W$-$F450W)$_{AB}$ vs F300W$_{AB}$ for the F300W-selected sample.  Asterisks
represent sources detected in F450W and F300W bands,
while circles mark sources undetected in the F450W band.}
\label{colori_U2}
\end{figure*}

Then we cross-correlated the catalogue obtained from the combined
image with F814W and F300W samples selected according to
our criteria (see Section 2): we ended up by having two sample
selections. Starting from each sample, we assigned a lower limit in
magnitude to sources undetected in any of other bands. The limiting
magnitude is the 5$\sigma$ isophotal magnitude within the isophote
measured in the combined image.  In Figures \ref{colori_I1},
\ref{colori_I2}, \ref{colori_U1}, \ref{colori_U2} the color-magnitude
diagrams of the two samples are shown together with the median
locus. The error bars are the standard deviation from the mean of the
values in each bin. The most notable feature is the initial trend
towards blue colors followed by a flattening in the last bins.

\begin{figure}
\centerline{\psfig{figure=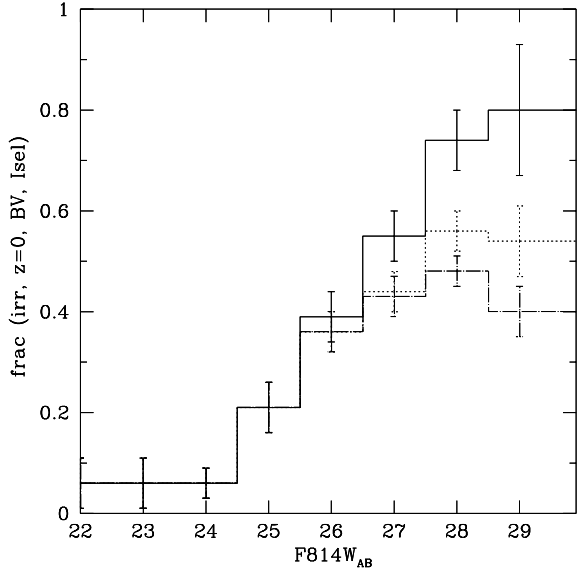,height=80mm}}
\caption{Fraction of sources in the F814W selected sample with
(F450W$-$F606W)$_{AB}$ bluer than a typical irregular galaxy (Coleman,
Wu, Weedman, 1980). The solid line refers to sources detected both in
F814W and F450W bands, the dashed line histogram
includes also lower limits to magnitudes for sources undetected in the
F450W band. The dot-dashed histogram includes also galaxies
undetected both in F606W and F450W bands, with lower
limits in both magnitudes.}
\label{frac_irr}
\end{figure}

We then compared our colors with colors measured from spectra by
Coleman, Wu \& Weedman (1980, CWW hereafter) after convolution
with throughput curves for the WFPC2 filters. In each magnitude bin,
we estimated the fraction of sources with (F450W$-$F606W)$_{AB}$ bluer
~than ~an ~irregular galaxy at $z=0$ (i.e. F450W$_{AB}-$
F606W$_{AB}<$0.18). As shown in Figure \ref{frac_irr}, at
F814W$_{AB}>27$ about $50\%$ of sources have (F450W$-$F606W)$_{AB}$
bluer than a typical local irregular.  For the F300W-selected
catalogue we also estimated the percentage of galaxies with
(F300W$-$F450W)$_{AB}$ bluer than a typical~irregular: at
F300W$_{AB}$=26 this percentage is almost $80\%$.

We will discuss these features in  Section 5.2. 

\section{Discussion}

\subsection{Number Counts}

Our results show a decreasing slope at redder wavelengths for the
whole sample, in good agreement with other works. The very~ faint~
end~ is flatter than 0.2 in F450W$_{AB}$, F606W$_{AB}$, F814W$_{AB}$,
while in F300W$_{AB}$ the slope is steeper. ~The ~difference ~with
~Pozzetti ~et al. F300W$_{AB}$ counts is likely due to a different
selection of the catalogue: they selected the sample~ in~ a~ red~ band
~(on~ a summed F606W+F814W image) to derive F300W-band counts and used
the isophotal magnitude instead of a \emph{pseudo-total} magnitude.
They ~assume ~that ~galaxies ~fainter ~than F606W$_{AB}<26$ have
approximately a (F300W$-$F606W)$_{AB}$ color not bluer than the
brighter galaxies, not biasing the red-selected sample against very
blue objects.
%discuss a potential bias against very blue objects due to the red-selection. These sources would affect U-band counts if they had U$-$V$\leq$0 at U$_{300}\approx$29 and U$-$V$\leq$0.5 at U$_{300}\approx$29.5. They analyze the bright part of the sample (F606W$_{AB}<26$) finding only 1$\%$ of the galaxies with U$-$V$\leq$0 and assume that fainter galaxies have approximately the same color. 

\begin{figure}
\centerline{\psfig{figure=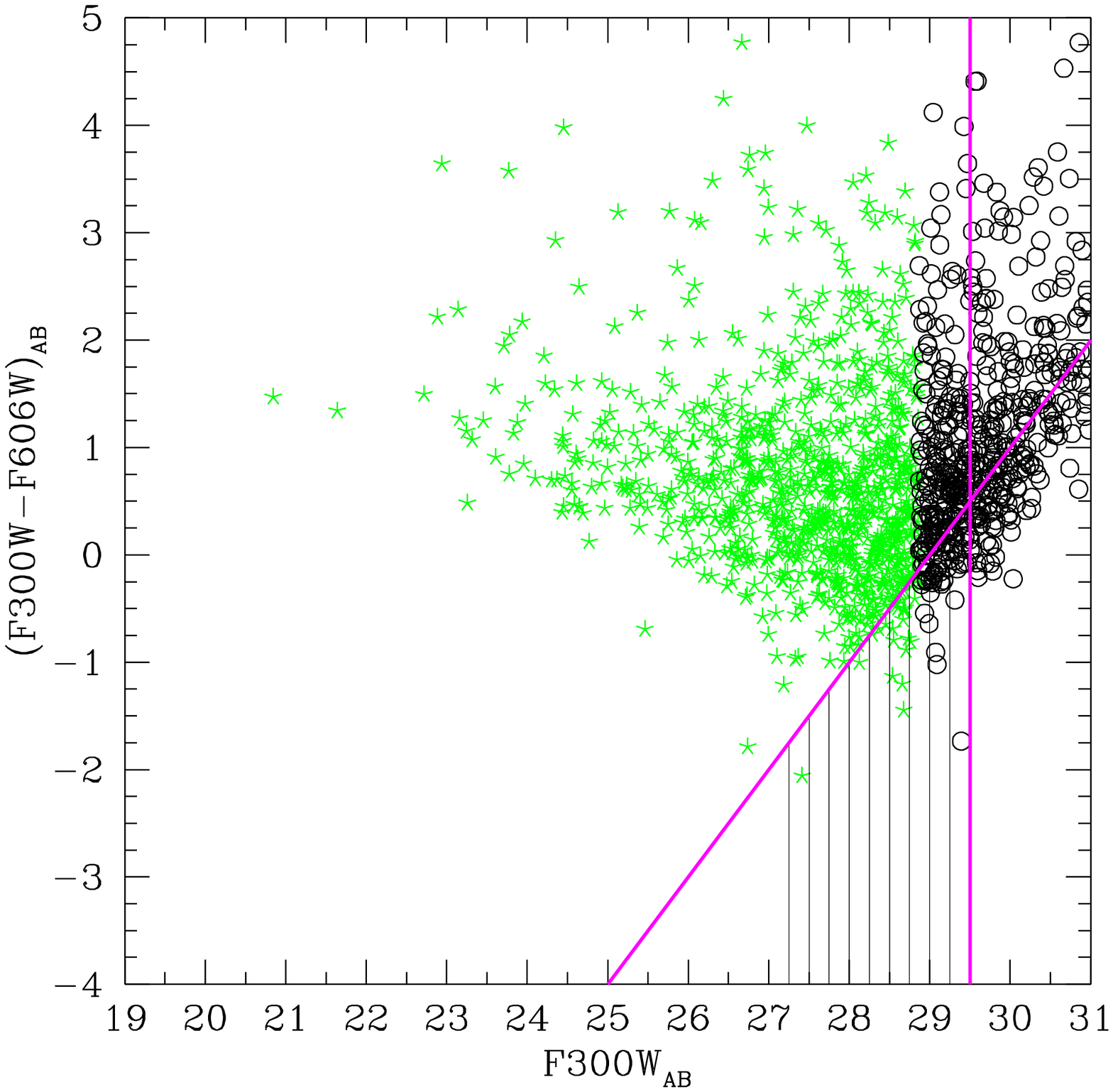,height=80mm}}
\caption{(F300W$-$F606W)$_{AB}$ color as a function of F300W$_{AB}$
isophotal magnitude for the F606W-selected sample. Circles represent galaxies with a lower limit in the F300W$_{AB}$ magnitude and consequently in the (F300W$-$F606W)$_{AB}$ color. The solid lines
represent the (F300W$-$F606W)$_{AB}$ color of a F606W$_{AB}$=29 galaxy
(F606W-band limiting magnitude considered by Pozzetti et al. 1998) and
the F300W$_{AB}$=29.5 limit. The apparent trend towards redder colors
for F300W$_{AB}>$28 is due to the F606W-band limiting
magnitude. }
\label{pozzU}
\end{figure}

\begin{figure}
\centerline{\psfig{figure=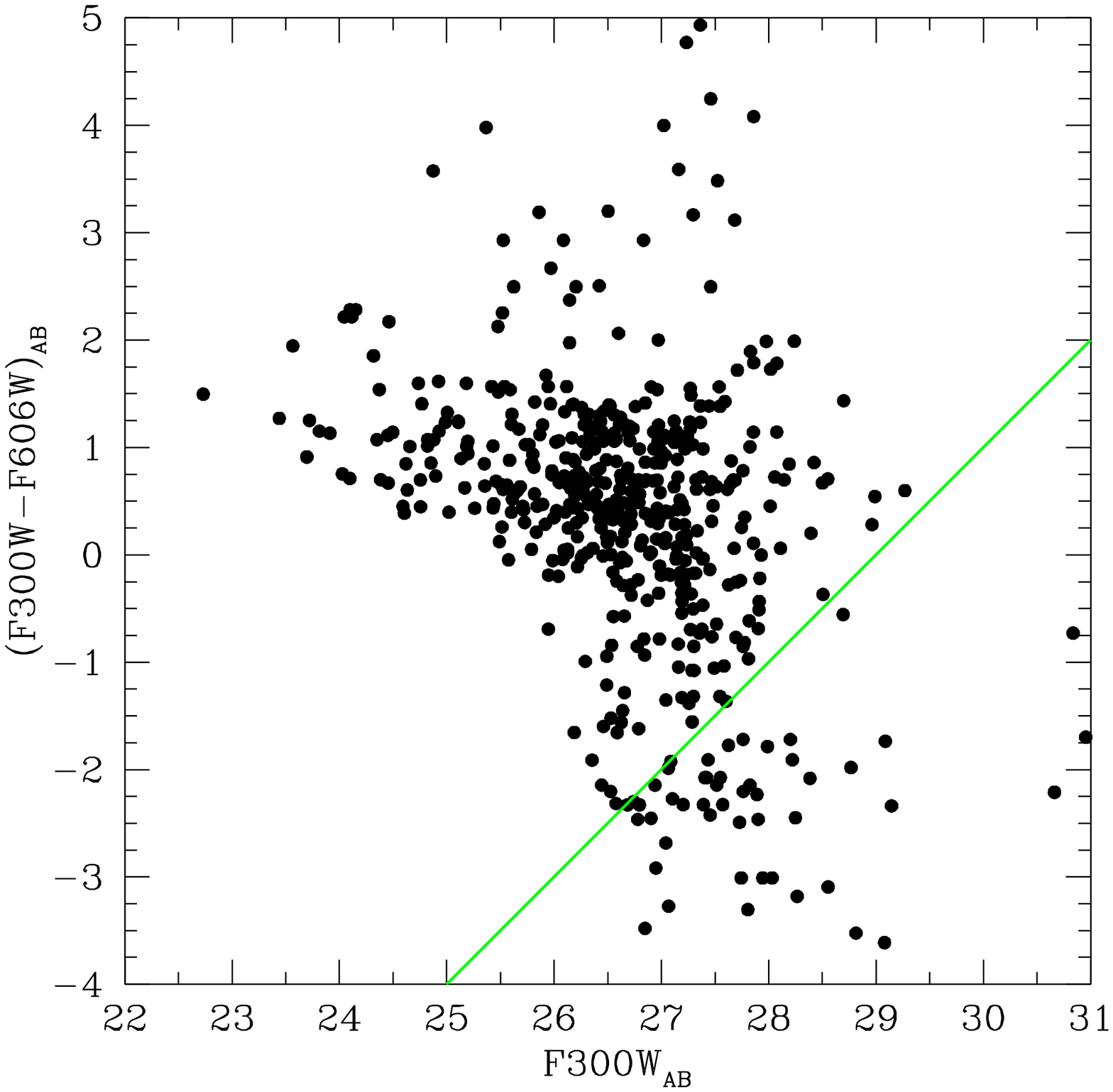,height=80mm}}
%\caption{U$-$V color as a function of F300W$_{AB}$ isophotal magnitude for the HDF-N galaxies . The solid lines represent the U$-$V color of a V$_{300}$=29 galaxy (V-band limiting magnitude considered by Pozzetti et al. 1998) and the F300W$_{AB}$=29.5 limit. }
\caption{(F300W$-$F606W)$_{AB}$ color as a function of F300W$_{AB}$
isophotal magnitude for the F300W-selected sample. The solid line
represent the (F300W$-$F606W)$_{AB}$ color of a F606W$_{AB}$=29 galaxy
(F606W-band limiting magnitude considered by Pozzetti et
al. 1998). }
\label{U}
\end{figure}

In Figure \ref{pozzU} we show the (F300W$-$F606W)$_{AB}$ color as a
function of F300W$_{AB}$ isophotal magnitude (as Pozzetti et al.) for
our F606W-selected sample, within the limiting F606W$_{AB}$
(i.e. $\approx30.2$) and for the F300W-selected one (Figure
\ref{U}). The colors for the F606W-selected sample are
obtained as described in Section 4. In order to obtain a sample with
characteristics similar to Pozzetti et al. sample, we considered for
F300W$_{AB}$ magnitude the flux within the isophote in the
\emph{metaimage}.

The ~~solid ~oblique ~line ~represents ~the ~(F300W $-$F606W)$_{AB}$
color of a F606W$_{AB}=29$ galaxy (the F606W-band limiting magnitude
of Pozzetti et al.): all galaxies in the hatched area would be not
counted since they are fainter than F606W$_{AB}>29$, the vertical line
is the F300W$_{AB}$=29.5 limit.  The F606W$_{AB}$ limiting magnitude
biases the sample against blue objects, that is the last bins are
affected by censored data (see next Section), thus making the
(F300W$-$F606W)$_{AB}$ color appear redder.
%We repeated the same analysis on the HDF-N catalogue released on December 23 1998 (http://www.stsci.edu/ftp/science/ hdfsouth/catalogs.html), regardless of the F606W$_{AB}$ limiting magnitude, and found a remarkable fraction of galaxies with  U$-$V$\leq$0 at U$_{300}\approx$29 and U$-$V$\leq$0.5 at U$_{300}\approx$29.5, that is galaxies in the area under the intersection of the two lines (corresponding to the U and V limiting magnitudes of Pozzetti et al.) in Figure 7. The F606W$_{AB}$ limiting magnitude biases the sample against blue objects, that is the last bins are affected  censored data (see next Section), thus making the U$-$V color appear redder. Relying on the robustness of the HDF-N catalogue in the faintest bins, but 
Considering the bluing trend observed until F300W$_{AB}$=29, and the
fact that the F606W$_{AB}$ limiting magnitude biases the sample
against blue objects, this implies that F300W-band counts recovered
from a red-selected sample are affected by this color bias and
consequently by incompleteness in the last bins. The flattening found
by other Authors may be mainly caused by the influence of such
incompleteness (noise being equally important for faint galaxies, be
them at low or high redshift) and not to the ``crossing'' of the Lyman
break.
%, contrary to what they argue from their fluctuation analysis. 
% The difference with their result may also be due to the different catalogue used, since their number counts are derived from version 2 of the HDF-N catalogue (February 29, 1996). 

%\begin{figure}
%\centerline{\psfig{figure=pozzU_HDFN.ps,height=80mm}}
%\caption{U$-$V color as a function of F300W$_{AB}$ isophotal magnitude for the HDF-N galaxies . The solid lines represent the U$-$V color of a V$_{300}$=29 galaxy (V-band limiting magnitude considered by Pozzetti et al. 1998) and the F300W$_{AB}$=29.5 limit. }
%\end{figure}

\subsection{Colors}

As explained in Section 4 we selected two samples: the F814W-selected
one should trace the global features of the HDF-S sample and the
F300W-selected sample evidencing star-forming galaxies at moderate
$z$.

 The F814W-selected catalogue has by far the biggest number of
sources, so that the number of galaxies with lower limits in the other
bands is rather high. We applied a linear fit to the color-magnitude
relation by making use of ``survival analysis'' (Avni et al. 1980,
Isobe et al. 1986), in order to avoid uncertainties introduced by the
heavy influence of censored data.  We applied a linear regression
based on Kaplan-Meier residuals. This analysis has suggested an almost
flat relation for (F450W$-$F814W)$_{AB}$ ~vs ~F814W$_{AB}$,
(F450W$-$F606W)$_{AB}$ ~vs ~F814W$_{AB}$ and ~(F606W$-$F814W)$_{AB}$
vs ~F814W$_{AB}$ for ~F814W$_{AB}>24$, while (F300W$-$F450W)$_{AB}$
~vs ~F814W$_{AB}$ did not reach convergence. Smail et al. (1995)
noticed a similar tendency on their sample limited at R=27.  After an
initial bluing the median V$-$R color gets redder, V$-$I gets flat,
while R$-$I keeps on following a bluing trend.

This trend may suggest the presence of a flat spectrum population,
whose colors seem to saturate but whose contribution is more and more
important at faint magnitudes, as confirmed by the rising fraction of
galaxies with (F450W$-$F606W)$_{AB}$ bluer than a typical irregular
galaxy.  Broadly speaking galaxy colors are dominated by blue sources,
as noted by Williams et al. (1996)in the HDF-N, that is the median
color is always bluer than typical local samples.  The F300W-selected
sample shows a very blue (F300W$-$F450W)$_{AB}$ color, though the
relation (F300W$-$F450W)$_{AB}$ vs F300W$_{AB}$ seems almost
flat. These two features suggest a considerable contribution by flat
spectrum sources.  Comparing the median (F300W $-$F450W)$_{AB}$ color
with Bruzual \& Charlot (1993) and CWW predicted colors, the main
contribution may be due to sources with $z>0.5$, while
(F450W$-$F814W)$_{AB}$, (F606W$-$F814W)$_{AB}$ and
(F450W$-$F606W)$_{AB}$ are compatible with those of local irregular
galaxies.

\section{Morphological Number Counts}

We performed a morphological analysis of the F814W-selected sample,
limited to F814W$_{AB}$=25 (Volonteri et al. 2000). We chose a
quantitative classification of galaxies morphology, following Abraham
et al. (1994, 1996).  For each galaxy we measured an asymmetry index
($A$) and a concentration index ($C_{abr}$). The former is determined
by rotating the galaxy by 180$^\circ$ and subtracting the resulting
image from the original one.  The asymmetry index is given by the sum
of absolute values of the pixels in the residual image, normalized by
the sum of the absolute value of the pixels in the original image and
corrected for the intrinsic asymmetry of the background.  The
concentration index is given by the ratio of fluxes in two isophotes,
based on the analysis of light profiles.

We determined number counts by splitting the sample in three bins:
~E/S0, ~spirals and ~irregular/peculiar/ interacting according to our
quantitative classification. In every bin the error is estimated as
explained in Section 3, considering only contributions due to
Poissonian and clustering fluctuation, since according to our
simulations the sample at F814W$_{AB}$=25 is complete.

\begin{figure}
\centerline{\psfig{figure=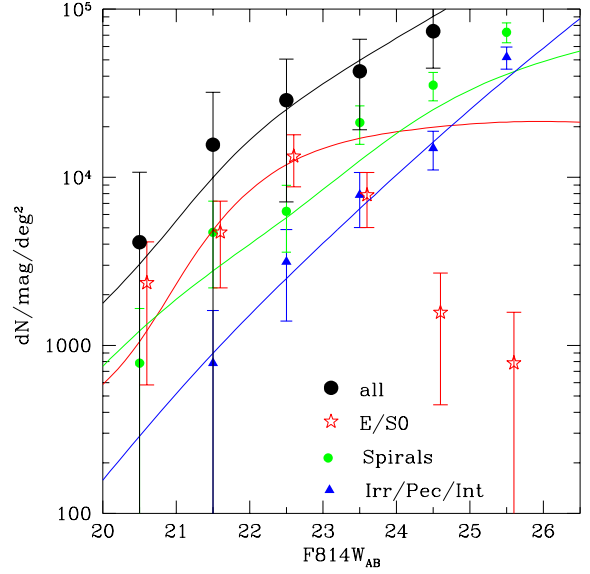,height=80mm}}
\caption{Morphological number counts: early type galaxies are shown as
circles, spiral galaxies are shown as pentagons and peculiar galaxies
as squares, the total counts are shown as filled circles. Late type
galaxies have steep slopes, $\gamma_{irr}=0.43\pm0.05$,
$\gamma_{spiral}=0.37\pm0.05$.  Lines show number counts model for a
flat cosmological model ($q_0=0.5$, $H_0=50$ km s$^{-1}$ Mpc$^{-1}$,
$\Lambda=0$) and the LF of Marzke et al. (1998): the models by types
are obtained by keeping the parameters which best fit the total
counts.}\label{morph_c}
\end{figure}

In Figure \ref{morph_c} the decrease in differential counts for
early-type galaxies at F814W$_{AB}>$22.5, and the steeper slope for
spiral galaxies and irregular/peculiar/interacting ones is clear. They
are described by a slope $\gamma_{irr}=0.43\pm0.05$,
$\gamma_{spiral}=0.37\pm0.05$, we discuss this feature further in
Section 7.

 Abraham et al. (1996) and Driver et al. (1998) studied the
morphological number counts for all galaxies with F814W$_{AB}<$26.0
detected in the HDF-N.  Our results are in agreement with theirs:
number counts are dominated by late type galaxies and early type
sources show a flat curve. Results for late type galaxies are
compatible with a strong evolution in number or with a non negligible
fraction of sources being at moderate redshift, as suggested also by
the analysis of colors. This result fits in Marzke et al. (1997)
findings about the LF of galaxies as a function of their color: the
slope of the LF faint end is steeper at low luminosities for the
bluest galaxies ($M^*_B\approx -19.4$, $\alpha<-1.7$ for galaxies
bluer than B$-$V=0.4).  Actually by integrating this LF, within
$z=0.5$, the number density of galaxies bluer than an Irregular galaxy
is compatible with the number found in our sample.  The flattening of
the curve for early type sources may be due to a decrease in density
at higher redshift of elliptical galaxies, but it is also expected by
the $\alpha>-1$ slope of the LF of early type galaxies (e.g.  Marzke
et al. 1998). Number counts models nevertheless show that an
$\alpha>-1$ LF does not account for such a steep decrease (Figure
\ref{morph_c}).

Brinchmann et al. (1998), Im et al. (1999) and Driver et al. (1998)
studied the redshift distribution for different morphological types in
deep surveys.  Brinchmann et al. (1998) analyzed 341 galaxies at $z<1$
from the Canada-France Redshift Survey (CFRS, Lilly et al. 1995,
limited at F814W$_{AB}$=22.5), and Autofib/Low-Dispersion Survey
Spectrograph (LDSS, Ellis et al. 1996, limited at $b_j=24$). They find
the distribution of elliptical galaxies peaked at $z\approx0.5$, while
spiral galaxies have a shallow distribution and the excess of peculiar
sources begins at $z\approx0.4$. According to their conclusions, the
number of regular galaxies is compatible with a passive evolution
model, while the excess of peculiar ones suggests an active luminosity
evolution and-or number evolution.  Im et al. (1999) for a sample of
464 galaxies limited at I$=21.5$ do agree with Brinchmann et
al. (1998). Driver et al. (1998) analyzed HDF-N sources, with
photometric redshifts, and underline the excess both in spiral
galaxies at $z=1.5$ and irregulars at $z>1$, suggesting number
evolution and passive evolution for dwarf LSBG at $z<1$ giving the
density of irregular galaxies at low redshift.

\section{Constraining the Redshift of Galaxies from Photometric Data}

In a deep survey, such as the HDF-S, the contribution of low redshift
foreground sources should be disentangled from the contribution of
high redshift ones. We first tried to split the two populations with
very simple considerations, regarding galaxy colors and counts.

\begin{figure}
\centerline{\psfig{figure=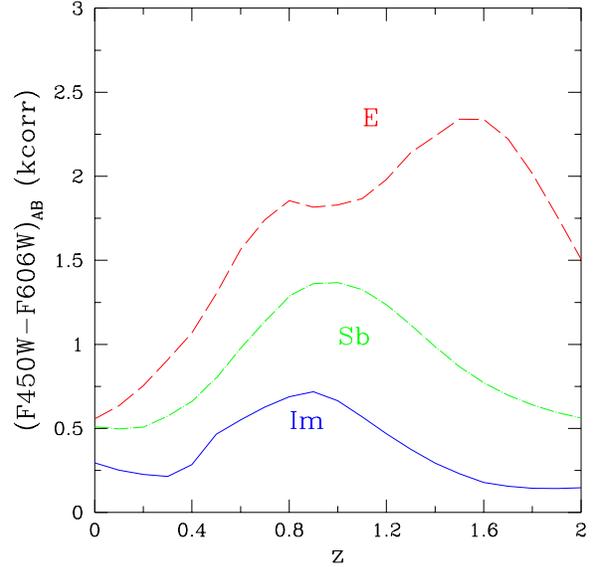,height=80mm}}
\caption{(F450W$-$F606W)$_{AB}$ color from Coleman, Wu e Weedman
(1980) spectra, convolved with HST filter responses. $K-corrections$
have been applied, but no evolution has been considered. The solid
line refers to irregular galaxies, the dot-dashed line to disk
galaxies, the dashed line to ellipticals.}  
\label{CWW}
\end{figure}

\begin{figure}
\centerline{\psfig{figure=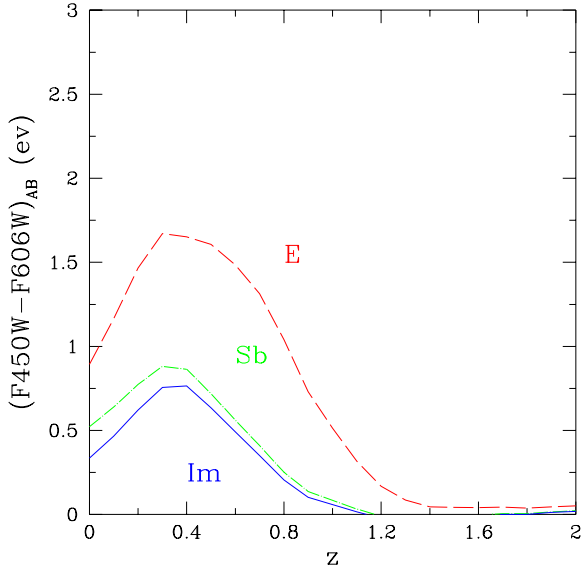,height=80mm}}
\caption{As in Figure \ref{CWW} but considering Bruzual \& Charlot
(1993) of synthetic stellar evolution spectra, with a
Miller-Scalo IMF. The star formation rate is exponentially decaying
with timescales 1 Gyrs and 5 Gyrs for E and Sb respectively, and is
constant for Im. }\label{BC}
\end{figure}

We compared colors of HDF-S galaxies with those predicted by CWW
spectra and~ by~ synthetic~ spectra~ of Bruzual \& Charlot (1993, BC
hereafter).  While BC account for evolution, the ~CWW ~spectra ~take
~into ~account $K-corrections$.  In Figures \ref{CWW}-\ref{BC}
apparent colors from CWW and BC are shown.  We limited our analysis at
$z<2$ to avoid relying too much on the UV part of spectral energy
distributions which has a strong influence at higher redshift.

By comparison with colors predicted by CWW or BC spectra, the observed
very blue galaxies (med(F450W$_{AB}-$ F606W$_{AB}$) = 0.06,
~med(F606W$_{AB}-$ F814W$_{AB}$) = 0.24) are compatible with two very
different ranges in redshift: $z<0.2$ or $z>1.5$ (and perhaps beyond
our analysis limit). The slope of morphological number counts is in
good agreement with this hypothesis, assuming that most of the faint
late type galaxies are at moderate redshift. In the hypothesis of the
moderate redshift population ($z\leq0.2)$ we computed number counts as
a function of optical colors, which should reflect rest-frame colors
for this range in z. We divided the F814W-selected sample in a ``red''
subsample, composed by sources redder than the median colour of the
sample (F450W$_{AB}-$ F606W$_{AB}$=0.45) and a ``blue'' subsample of
the remaining sources. We limited our analysis at F814W$_{AB}$=26,
where the catalogue is complete, according to our simulations.

\begin{figure}
\centerline{\psfig{figure=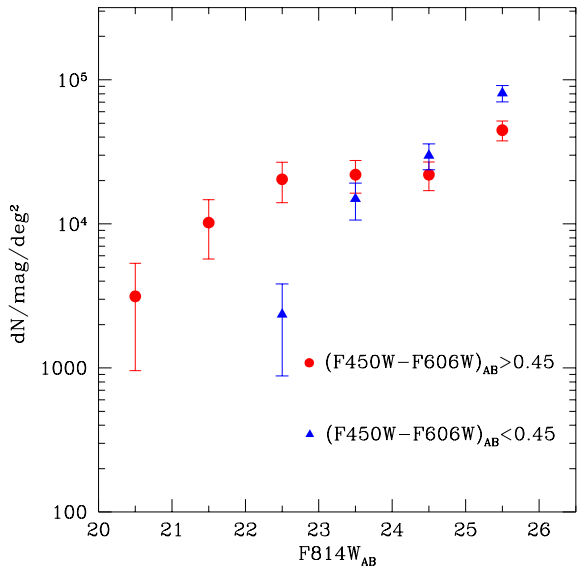,height=80mm}}
\caption{Number counts for blue and red galaxies
(i.e. (F450W$-$F606W)$_{AB}<0.45$ or vice versa). The blue sources
have a very steep slope ($\gamma_{blue}\sim0.49\pm0.01$), while red
sources have a shallower end.  }
\label{col_c}
\end{figure}

The curve for blue sources is much steeper than the red one, as shown
in Figure \ref{col_c}. We estimated $\gamma_{blue}\sim0.49\pm0.01$.
This value is therefore compatible with the previous hypothesis, that
a non negligible fraction of the blue galaxies may have a moderate
redshift, contributing with an almost Euclidean slope to the counts or
with number evolution, with very faint galaxies merging at high
redshift.  However we cannot rule out the hypothesis that such a steep slope can be due to earlier-type evolving galaxies which move into the blue sample at high redshift.  In such a case the steep counts described by the faint blue sample would not require a significant rate of merging to be justified.

In order to check the low-redshift solution we compared the
characteristics of those galaxies to the predictions of a local
luminosity function.  The brightest of the very blue sources would
have $M_B\geq -18$ if placed at $z=0.2$ ~(assuming ~$H_0$=50 km
s$^{-1}$ Mpc$^{-1}$ ~~and a (F450W$-$F814W)$_{AB}$=1).

Considering the local luminosity function (LF) of blue galaxies as
estimated by Marzke et al. (1997), described by $\alpha=-1.8$ and
$M^*_B\approx-$19.5 and integrating within $z=0.2$ and $M_B=-$18, the
number density found is compatible with all very blue galaxies
(i.e. galaxies with (F450W$-$F606W)$_{AB}$ and
(F606W$-$F814W)$_{AB}$ bluer than a local Irregular galaxy)
being at such a low redshift.  The first raw technique used to
constrain the redshift of sources suggests that a high fraction of
faint galaxy is composed by low redshift ($z<0.2$) sources,
nevertheless the ``high redshift solution'' is not ruled out, since
steep counts may be due also to merging and our color analysis is very
naive.

We then made our analysis better by means of the photometric redshift
technique. We computed photometric redshifts for the galaxies in the
F814W catalogue by means of the public code {\it hyperz\/}
(Bolzonella et al. 2000).  The method basically consists in a
comparison between the observed magnitudes and the photometry expected
from a set of Spectral Energy Distributions (SEDs).  The efficiency of
the method relies on the presence of strong spectral features, in
particular the $4000$\,\AA ~ break and the Lyman break for galaxies at
high redshift.  To reduce the uncertainties estimating $z_{phot}$, the
set of filters must be able to identify these characteristics,
spanning a wide range of wavelength.  The photometric redshift
$z_{phot}$ is computed by a $\chi^2$ minimization, considering the
whole set of possibilities with different combinations of the involved
parameters.

In this procedure we decided to use the SEDs built from the Bruzual \&
Charlot's synthetic library (GISSEL98, Bruzual \& Charlot 1993),
rather than the observed mean local spectra by CWW, because tests on
HDF-N indicate a slightly lower dispersion at $z<2$. In any case, CWW
spectra lead to not very different conclusions.  Hence, we selected 5
spectral types, characterized by different star formation rates,
matching the observed colors of the morphological galaxy sequence; a
subsample of ages allowed by the GISSEL library is also considered.
Moreover, we took into account the possible presence of dust applying
the reddening law by Calzetti et al. (2000) for different values of
$E_{B-V}$.  The flux decrement produced by intervening neutral
hydrogen is computed following the recipe by Madau (1995).  Only
spectra with solar metallicity are considered here, because the
metallicity can be regarded as a secondary parameter.

The accuracy of photometric redshift estimate obtained from the
considered set of filters can be studied by means of simulations on
synthetic catalogue.

Some degeneracy can be found (Bolzonella et al., 2000), with a non
negligible part of galaxies lying at $z\simeq 1$ -- $2.4$ incorrectly
located at low redshift.  These galaxies show frequently a probability
function with several comparable peaks, at low and high $z$, due to
the degeneracy in the parameter space.  Near infrared data can in
principle avoid these uncertainties.  Nonetheless, also with the
available set of filters, we can guess that the objects with
$z_{phot}\gtrsim 1$ really belong to this redshift, because high-$z$
objects are rarely misidentified as low-$z$ galaxies.

Photometric redshifts for our catalogue indicate that all very blue
galaxies are at $z_{phot}\geq1$ (med($z_{phot}$)=1.6). This feature,
along with their steep slope (see Section 5), suggests a high merging
rate for galaxies at $z_{phot}\geq1$. Vice versa very red sources have
a shallower redshift distribution (med($z_{phot}$)=0.6), though at
F814W$_{AB}>25$ most sources have a high redshift.

\begin{figure}
\centerline{\psfig{figure=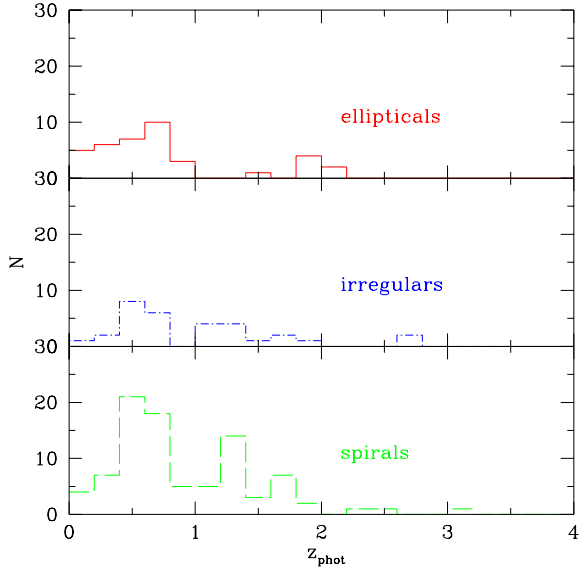,height=80mm}}
\caption{Morphological redshift distributions for
F814W$_{AB}<$25. }
\label{z_histo}
\end{figure}

Similarly, at F814W$_{AB}<$25, late type galaxies have a higher median
redshift (med($z_{phot}$)=1.1) than ~early ~type ~ones
(med($z_{phot}$)=0.6)~and the redshift distribution (Figure
\ref{z_histo}) is similar to Driver et al.  (1998) one, though the
number of sources at $z_{phot}>1.5$ is lower.  There is a point to be
emphasized about redshift distributions and morphological types: the
detection of galaxies at high redshift is seriously biased against
elliptical galaxies because of $K-corrections$ in the F814W-band.  At
$z\approx1$, early type galaxies have $K-corrections$ ($\approx 1$
mag) greater than late type $K-corrections$, making them fainter.  It
means that a cut in apparent magnitude, such as F814W$_{AB}<$25,
biases the sample toward galaxies with lower $K-corrections$.  To skip
this problem, we tried to select a volume-limited sample, based on our
photometric redshifts. When selecting all galaxies with
$z_{phot}\leq1$, we found only 15$\%$ to be with F814W$_{AB}<$25, that
is with a reliable morphological classification according to our
method.  Vice versa it means that a cut off in apparent magnitude does
not offer a fair sample for a redshift distribution of the different
morphological types.

Probably also the apparent decrease in early-type galaxies number
counts, as shown in Figure \ref{morph_c}, is due to this bias.

\begin{figure}
\centerline{\psfig{figure=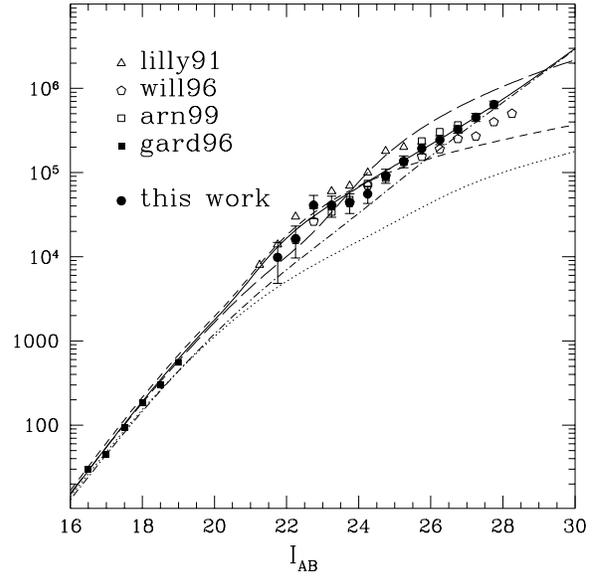,height=80mm}}
\caption{Number counts models compared to various counts obtained in
literature and in this work. All models adopt Marzke et al. (1998)
luminosity function, the dot-dashed line refers to three Schechter
fits for elliptical, spiral and irregular galaxies with no evolution,
the solid line uses the same LFs with luminosity evolution and a
moderate merging, the dotted line is obtained by using a single
Schechter fit without evolution, the short dashed considers luminosity
evolution and the long dashed both luminosity evolution and strong merging
for the same LF.}
\label{ncmod}
\end{figure}

We compared our number counts with different models, by using galaxy
counts models by Gardner (1998). We adopted a flat cosmological model
($q_0=0.5$, $H_0=50$ km s$^{-1}$ Mpc$^{-1}$) in all models. We used
the luminosity function of Marzke et al. (1998) both considering a
single Schechter function for all morphological types and three
different Schechter fits for elliptical, spiral and irregular
galaxies.  Our counts are best fitted by considering three LFs,
luminosity evolution and a moderate merging (Figure \ref{ncmod})
consistent with our interpretations. The merging rate in Gardner
(1998) follows Rocca-Volmerange \& Guiderdoni (1990) with number
evolution parameterized as $\phi^* \propto (1+z)^{\eta}$ in the
Schechter fit to the LF, and in order to conserve the luminosity
density, $L^* \propto (1+z)^{-{\eta}}$; $\eta$ is a free parameter. 
Good fits to our data are obtained  with $\eta\leq0.5$ in the case of three LFs for the cosmological parameters considered.

\section{Summary and Conclusions}

The HDF-S represents a unique opportunity for the study of ~faint
~galaxies ~up to~ now, ~both ~for ~its ~depth
(F814W$_{AB,lim}\approx29$ for detection and F814W$_{AB,lim}\approx27$
for completeness) and spatial resolution (FWHM$\approx$0.2 arcsec). We
presented here colors and number counts of HDF-S galaxies, along with
number counts determined by splitting the sample considering the
morphology and the colors of the galaxies. We also analyzed the
photometric data to constrain the redshift of HDF-S galaxies, and
determine the contribution of different redshift populations to the
counts.  The main results are the following:
\begin{itemize}
\item the number-counts relation has an increasing slope up to the
limits of the survey in all four bands. The slope is steeper at
shortest wavelengths;
\item the number counts model which best fits our data is obtained by
considering three different Schechter functions for elliptical, spiral
and irregular galaxies (Marzke et al. 1998), luminosity evolution and
a moderate merging ($\eta < 0.5$);
\item optical colors show that the sample contains a high fraction of
galaxies bluer than local sources, for instance at F814W$_{AB}>27$
about $50\%$ of sources have (F450W$-$F606W)$_{AB}$ bluer than a
typical local irregular galaxy;
\item after an initial blueing trend the color-magnitude relation gets
flat, suggesting a color-saturation due to a strong contribution of a
flat spectrum population in the faintest bins;
\item morphological number counts ( F814W$_{AB}<25$) are dominated by
late type galaxies ($\gamma_{irr}=0.43\pm0.03$, $\gamma_{spir}=
0.37\pm0.03$), while early type galaxies show a steep decrease in the
faintest bin;
\item photometric redshifts of our sample galaxies show that the
galaxies contributing with a steep slope to the number counts have
$z_{phot}\gtrsim1$, suggesting a moderate merging;
\item morphological redshift distributions and number counts are
biased against elliptical galaxies when using an apparent magnitude
cut-off.
\end{itemize}
%Deep number counts are a powerful tool for studying the evolution of galaxies, though the analysis of faint galaxies is difficult due to observational biases.
Our number counts are in good agreement with previous results, except
for the F300W-band number counts in the HDF-N.
% Our simulations regarding incompleteness signal that for U$_{300}>$26.5 the U-detection is incomplete (we stress that the correction in the last bin $26.5<$ U$_{300}<$27 may not be robust). On the other hand, we show that also a selection in a different, deeper band of a multi-band survey is incomplete if selection effects are not counted. For instance a selection in the F606W$_{AB}$ band  is affected by the V-detection incompleteness itself (at F606W$_{AB}>28$) and a color bias (at U$_{300}>27$). 

 We do not see a flattening in F300W-band number counts, limiting our
analysis at F300W$_{AB}<$26.5.
%Moreover considering the high degree of incompleteness found ($\approx80\%$), the flattening found by other Authors may be mainly caused by the influence of such incompleteness (noise being equally important for faint galaxies, be them at low or high redshift) and not to the ``crossing'' of the Lyman break.
 A steep F300W-slope may be explained both with a high fraction of low
redshift galaxies or with merging or a mixture of them. In the former
case, intrinsically faint, low redshift galaxies, imply a steep faint
end slope of the LF in the F300W-band ($\alpha\approx-2$) in the
latter case we emphasize that if merging is the reason of the steep
F300W-band counts, it would take place at $z<2$ since the Lyman break
prevents us from detecting galaxies in the F300W-band at higher
redshift. A way of detecting merging and hence to check the validity
of the hierarchical model, could be in principle the study of the
evolution with look-back time of some structural parameters of
galaxies, such as their intrinsic size.

Morphological number counts are in agreement with previous results, as
well as the morphological redshift distributions. The morphological
classification is, however, possible only for bright galaxies. The cut
in apparent magnitude biases the sample against early-type galaxies,
due to their large $K-corrections$. A correct test implying
morphological classification should be done using K-band data (an
extension of the test proposed by Kauffmann \& Charlot, 1998) or a
volume-limited sample, with morphology determined from
spectra. Extracting a volume-limited sample from a magnitude limited
survey does not fit the goal since at $z\lesssim1$ almost 80$\%$ of
the galaxies are too faint for a reliable morphological classification
from imaging.

\begin{acknowledgements}
We would like to thank Roser Pell\'o for the code {\it hyperz\/}, R.
Abraham and J.P.  Gardner for making available their software and
A. Buzzoni, M. Massarotti for interesting and stimulating discussions.
\end{acknowledgements}

\end{document}